\documentclass[journal,comsoc]{IEEEtran}
\ifCLASSINFOpdf
\else
\fi
\usepackage[T1]{fontenc}
\usepackage{cite}
\usepackage{graphicx}
\usepackage{epstopdf}
\usepackage{amsmath}
\usepackage{algorithm}
\usepackage{algpseudocode}
\usepackage{tabularx}
\usepackage{dblfloatfix}
\usepackage{caption}
\usepackage{subcaption}
\usepackage{multirow}

\begin{document}
%
\title{Novel Sparse-Coded Ambient Backscatter \\Communication for Massive IoT Connectivity}
%
%
%

\author{Tae Yeong Kim and Dong In Kim
\thanks{T. Y. Kim and D. I. Kim are with the School
of Information and Communication Engineering, Sungkyunkwan University (SKKU), Suwon 16419, Korea (e-mail: \{tangoeye,$\,$dongin\}@skku.edu).}}

\maketitle

\begin{abstract}
Low-power ambient backscatter communication (AmBC) relying on radio-frequency (RF) energy harvesting is an energy-efficient solution for batteryless Internet of things (IoT). However, ambient backscatter signals are severely faded by dyadic backscatter channel (DBC), limiting connectivity in conventional orthogonal time-division-based AmBC (TD-AmBC). 
In order to support massive connectivity in AmBC, we propose sparse-coded AmBC (SC-AmBC) based on non-orthogonal signaling. 
Sparse code utilizes inherent sparsity of AmBC where power supplies of RF tags rely on ambient RF energy harvesting. 
Consequently, sparse-coded backscatter modulation algorithm (SC-BMA) can enable non-orthogonal multiple access (NOMA) as well as $M$-ary modulation for concurrent backscatter transmissions, providing additional diversity gain. 
These sparse codewords from multiple tags can be efficiently detected at access point (AP) using iterative message passing algorithm (MPA). 
To overcome DBC along with intersymbol interference (ISI), we propose dyadic channel estimation algorithm (D-CEA) and dyadic MPA (D-MPA) exploiting weighted-sum of the ISI for information exchange in factor graph.
Simulation results validate the potential of the SC-AmBC in terms of connectivity, detection performance and sum throughput.
\end{abstract}


%
\IEEEpeerreviewmaketitle

\section{Introduction}
%
%
%
%


\IEEEPARstart{M}{assive} connectivity is a key component of future Internet of things (IoT) where radio frequency (RF) devices continuously interact with humans' daily routine.
The vision includes smart home gadgets connected to Wi-Fi (Wireless Fidelity) \cite{passive,backfi}, real-time interactive applications such as Bluetooth exchanging video/audio data with smart phone \cite{ble}, and inventory management in huge warehouse \cite{drone}. 
In these applications, RF devices should be designed in low power for prolonging their lifetime and small-form factor for saving cost. 
Unfortunately, it is reported that most of the power in RF devices are consumed by oscillator which generates sinusoidal carrier signals for modulation of data. The high power consumption for this digital-to-analog converter (DAC) hinders connectivity for numerous IoT applications with energy-constrained RF devices \cite{ambB,wifiB,turbo}. 
Hence, alternative modulation without the oscillator is considered to be a promising solution for fulfilling IoT requirements.

Ambient backscatter communication (AmBC) is an innovative solution for batteryless IoT, since it replaces power-hungry oscillator with passive load impedances connected to RF antenna \cite{ambB,wifiB,turbo}. 
RF tags, sort of RF devices composed of low-power micro-controller and load impedances, can reuse ambient signals such as television (TV) \cite{ambB,turbo} and Wi-Fi \cite{passive,backfi,wifiB} for sending their data.
Instead of generating RF sinusoidal signals, RF tags simply reflect incoming ambient signals and modulate their data by switching load impedances.
These load impedances implemented by passive circuit components, resistors and capacitors can constitute 2-dimensional signal constellation space spanned by real and imaginary axes of reflection coefficients \cite{16qam,qam}. Only a few $\mu\mathrm{W}$ of power is required for AmBC, which is in sharp contrast to traditional active RF radios demanding few tens of $\mathrm{mW}$. 
In addition, ultra-low power consumption of backscatter tags can sustain with ambient RF energy harvesting (EH). Ambient RF signals can be recycled not only for EH but also for information transmission over RF signals by controlling reflection coefficients at RF tags. Therefore, AmBC has strong advantages in terms of energy efficiency and installation cost by reducing size of RF hardware. 

\subsection{Related Works}
Despite of its benefits, AmBC has several challenges which are originated from inherent characteristic of RF signals and channels.
Firstly, the ambient signals are not dedicated to backscatter communications. Hence, some of backscatter schemes based on unmodulated excitation signals such as energy beamforming \cite{retro,RelayB}, space-time coding \cite{dyadic5, dyadic6}, frequency up/down conversion in radio-frequency identification (RFID) \cite{ble, drone} may not be available in AmBC. 
Without dedicated RF infrastructure (i.e., 915 MHz RFID reader and power beacon), previous works such as LoRa \cite{LoRaB}, rateless code \cite{rateless}, statistical inference \cite{inference}, time-hopping spreading spectrum \cite{TH} are not directly applied to AmBC scenarios. 
Since low-rate symbols are modulated on high-rate ambient signals, modulations and detections should be redesigned for applications of AmBC.
Second, channel propagation of AmBC is distinguished from that of active radios \cite{dyadic1,dyadic2,dyadic3,dyadic4,dyadic5,dyadic6}. 
Compared to one-way channel between transmitter and receiver \cite{wireless}, backscatter radios experience two-way channel \footnote{Here we assume {\em monostatic} ambient backscatter where the reader (e.g., Wi-Fi AP) generates RF signals intended for legacy (Wi-Fi) users (instead of RF tags) and also acts as the receiver for the signals reflected from RF tags \cite{backfi,ofdm2}. On the other hand, there is {\em bistatic} ambient backscatter where the RF signals are generated from the separated TV or FM station, and cellular base station, while the reader receiving the reflected signals from RF tags \cite{ambB, turbo}.}: forward channel from reader to tag; backward channel from tag to reader. The cascade of two channels is referred to as {\em dyadic backscatter channel (DBC)} and exhibits significant channel attenuations \cite{dyadic1} as well as delay spreads \cite{ofdm2} compared to the one-way channel. Eventually, the phenomena leads to low signal-to-noise ratio (SNR) for detection and severe intersymbol interference (ISI).

To overcome the inherent limitations in AmBC, various modulation and detection techniques based on diversity combining schemes are proposed.
Multiple receive antennas can enhance detection performance in AmBC. For example, interference cancellation schemes using channel differences \cite{turbo}, or zero-forcing \cite{cooperative} are proposed. 
The multi-antenna configurations are useful for achieving spatial diversity gain and effective as the number of receive antennas increases.
On the other hand, modulation and detection based on spreading sequences can also improve reliability achieving temporal diversity gain.
In \cite{backfi, ambB}, detection SNR of AmBC is reinforced using maximal-ratio combining of backscatter samples.
In \cite{ofdm1}, periodicity of ambient signal is utilized using cyclic prefix (CP) in orthogonal frequency division multiplexing (OFDM) sequences. Orthogonal spreading structure similar to code-division multiple access (CDMA) is proposed in \cite{turbo}.

However, most of AmBC assumed orthogonality in both modulations and detections. For instance, non-overlapping backscatter symbols are typically represented on 2-dimensional Euclidean space (e.g., on-off keying (OOK) \cite{ambB, wifiB}, phase-shift keying (PSK) \cite{backfi}, ternary encoding \cite{codingB}) with orthogonal multiple access (OMA) such as time-division multiple access (TDMA) \cite{ambB-TD} for providing connectivity. For detections, conventional AmBC ignored ISI in DBC by assuming long period of backscatter symbol \cite{ofdm2} or simply filtered it, which reduces SNR for signal detections \cite{backfi, ofdm2}.
In AmBC, tag backscatters data in active mode and harvests RF energy in idle mode whose time portion is much larger than the active mode \cite{rateless}. Consequently, we can utilize the duty-cycling operation by employing {\em sparse coding} over ambient RF signals and support non-orthogonal multiple access (NOMA) for massive connectivity. In addition, low-rate backscatter data are delivered over high-rate ambient RF signal, which can be equivalently regarded as {\em spreading sequences} to provide diversity gain for reliable backscatter communications \cite{ofdm1}. Based on these observations, we can design non-orthogonal sparse coding and detector to realize massive connectivity for AmBC. 

\subsection{Proposed Sparse-Coded AmBC (SC-AmBC)}
On the other hand, non-orthogonal signaling in both modulation and detection is proposed recently. 
For example, sparse code multiple access (SCMA) proposed in Huawei company proved effectiveness in highly-overloaded network \cite{scma1,scma2}. Especially, sparse coding has three desirable properties which are suitable for massive connectivity:
\begin{itemize}
\item {\bf Grant-Free Uplink Transmission:} With sparse coding, RF tag can transmit data in the uplink based on predefined sparse codebook. The codebook structure can accommodate massive and bursty uplink traffic without request-grant procedure \cite{grant-free}. The property can lead to low-latency without scheduling and resource allocation in contrast to the cognitive radio based AmBC \cite{ambB-TD}.
\item {\bf NOMA:} Sparse code encodes data in multi-dimensional complex codewords providing higher degree of freedoms than conventional 2-dimensional AmBC modulators \cite{ambB, wifiB, backfi, codingB}. As a result, sparse code can support concurrent RF transmissions and multiple-access interference (MAI) can be efficiently resolved by iterative message passing algorithm (MPA) \cite{scma1}.
\item {\bf Projected $M$-ary Modulation:} In conventional AmBC modulators, $M$-ary modulation is only implemented by $M$ distinct load impedances increasing form factors of RF devices for higher-order modulations \cite{backfi, 16qam, qam}. On the other hand, with sparse coding, $M$-ary data can be represented by few constellation symbols using codeword projection/expansion method \cite{iter-Mary,code-project}. Thus, $M$-ary modulation can be realized in AmBC without increasing form factors of RF devices.
\end{itemize}
In addition, due to increasing network densification and delay spreads in multipath propagation, exploiting ISI can achieve better spectral properties than conventional detectors by removing CP \cite{FBMC}. Based on these observations, we propose sparse-coded AmBC (SC-AmBC) including design of modulator and detector for DBC.  
Our main contributions can be summarized as follows:
\begin{itemize}
\item Sparse-coded backscatter modulation algorithm (SC-BMA) supporting $M$-ary modulation for AmBC with small-form factors.
\item Dyadic channel estimation algorithm (D-CEA) fully capturing characteristics of DBC.
\item Dyadic MPA (D-MPA) based on dyadic factor graph exploiting the weighted-sum of ISI caused by DBC in information exchange process.
\end{itemize}

To the best of the authors' knowledge, implementing the sparse code for energy-constrained AmBC is not fully explored in literature.
The rest of the paper is organized as follows. Section \ref{sec:system} describes SC-AmBC and motivating examples of our model therein. Section \ref{sec:enc} presents the load modulation including sparse coding algorithm and channel estimation algorithm, and iterative detector based on DBC is proposed in Section \ref{sec:dec} . Simulation results are provided in Section \ref{sec:simulation}. Finally, Section \ref{sec:conclusion} concludes the paper.

\begin{figure*}
\centering
\includegraphics[width=13cm]{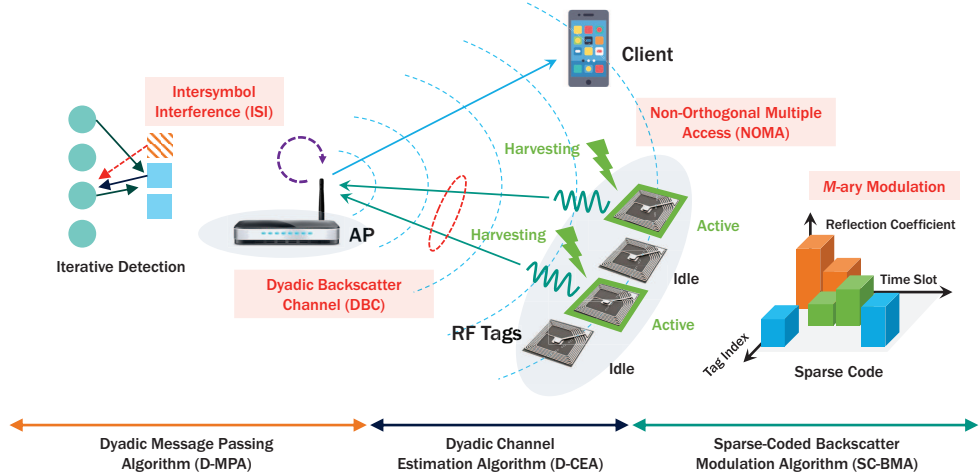}
\caption{Proposed AmBC network with single AP and multiple RF tags employing ambient RF energy harvesting and backscatter communications using sparse code.}
\label{fig:system}
\end{figure*}  

\section{System Model}\label{sec:system}
\subsection{Notation}
Matrices (vectors) are denoted with upper (lower) case bold-face letters (e.g. $\mathbf{A}$ or $\mathbf{a}$); the operator $\left ( \cdot  \right )^{H}$ denotes the Hermitian;  the operator $\left \| \cdot \right \|$ denotes the vector norm; the operator $\otimes$ denotes the Kronecker product; the set of real, complex and binary numbers are respectively denoted by $\mathbb{R}$, $\mathbb{C}$ and $\mathbb{B}$; the vector $\mathbf{1}_k$ denotes all-one column vector with size $k$ and the vector $\mathbf{0}_k$ denotes all-zero column vector with size $k$; 
the operator $p \left ( \mathcal{A}  \right )$ denotes the probability of the event $\mathcal{A}$.  the operator $\mathbb{E}\left ( \cdot  \right )$ denotes statistical expectation; $\mathcal{CN}(\mu,\sigma^2)$ denotes the circularly symmetric complex normal distribution with mean $\mu$ and variance $\sigma^2$.

\subsection{System Model}
We consider the AmBC network \cite{ABCN-kaibin} based on access point (AP) such as Wi-Fi as depicted in Fig.~\ref{fig:system}. 
The AP transmits data to its client as well as energy to nearby RF tags \cite{16qam,qam,codedqam, backfi}, replacing expensive and high-power RFID reader.
In terms of deployment cost, utilizing ambient signals is very attractive for backscatter communication as they exist everywhere for providing low-cost Internet connectivity (e.g., TV signal, Wi-Fi, BlueTooth). 
In AP's coverage area, there are multiple RF tags utilizing the ambient signals transmitted from the AP to downlink (DL) client. 
These RF tags can use the ambient signals for RF energy harvesting using rectifier circuit exempting large-sized battery.
If they harvest sufficient energy for activating backscatter communication, they modulate load impedances connected to micro-controller to transmit $M$-ary data instead of large and power-intensive oscillator component \cite{16qam,qam,codedqam, backfi}.
Depending on ambient energy harvesting, tags are only activated in small fraction of time and mostly remain idle to harvest sufficient energy.
The sparsity of ambient backscatter signals motivates us encoding data to multi-dimensional complex codewords \cite{scma1, scma2, scma3, dyadic3} selected by time-variant reflection coefficients \cite{hybridB, dyadic3} and enables low-complexity decoding even for massive number of concurrent RF transmissions and higher-order modulation.
Consequently, SC-BMA among RF tags can enable NOMA, while providing massive connectivity for low-cost RF tags.

Upon the tags' modulation, AP receives low-rate uplink (UL) backscatter signals from the tags while transmitting high-rate DL signals to the client. Assuming single antenna at AP where transmission and reception occur simultaneously in full-duplex, these signals are overlapped causing self-interference (SI) \cite{backfi, SIC}. 
In addition, high-sampling rate of DL signals significantly reduces sampling period of UL backscatter signals, causing ISI in multipath propagation environments \cite{ofdm2}.
Moreover, non-orthogonality of sparse code may cause MAI among active tags, degrading detection performance of concurrent RF transmissions \cite{scma1, scma2, scma3, scma4, log-MAP, LDS}. 
Therefore, the detection of backscatter signals requires mitigation of the three interferences (i.e., SI, ISI and MAI). Based on self-interference cancellation (SIC) \cite{SIC}, strong DL interference can be efficiently removed so that UL backscatter detection can be feasible in practice \cite{backfi}.
Then, D-CEA followed by low-complexity iterative D-MPA is newly implemented to effectively detect data in the presence of the ISI and MAI.
Clearly, the ambient backscatter rather exploiting interferences significantly improves spectral efficiency in three-folds: full-duplex \cite{backfi, SIC}; NOMA with sparse code \cite{scma1, scma2, scma3}; multipath-diversity combining \cite{wireless}. 

\begin{figure}
\centering
\includegraphics[width=8.9 cm]{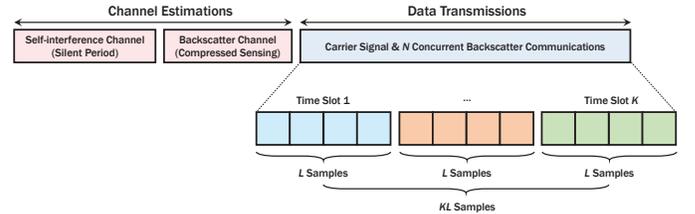}
\caption{Link layer protocol for AmBC with sparse code.}
\label{fig:protocol}
\end{figure}  

To support multiple RF tags, link layer protocol of AmBC is composed of channel estimation and data transmission phase as described in Fig.~\ref{fig:protocol}. In channel estimation phase, AP estimates the SI channel using silent period for RF tags \cite{backfi}. Then, the tags transmit sparse pseudo random symbols known to AP, and it can efficiently estimate backscatter channels using compressed sensing algorithms \cite{rateless, CS}. 
Moreover, as the DBC \cite{dyadic1,dyadic2,dyadic3,dyadic4,dyadic5,dyadic6} experiences two-way fading in contrast to the conventional one-way fading models, D-CEA capturing the characteristics of these channels will be described in Section \ref{sec:enc}. 

In data transmission phase, ambient signal from AP can be sliced into $K$ time slots, each with $L$ samples, to spread $N \geq K$ backscatter signals to increase signal-to-noise ratio (SNR) and reduce circuit-power consumption \cite{backfi, ambB, wifiB, turbo, ambB-analysis, ambB-noncoherent, codingB}. As RF tags employ predefined codebooks for data transmissions, active tags can immediately transmit data with low latency without resource allocation and scheduling procedures \cite{grant-free}.
At time slot $k\in \{1,2,\cdots,K\}$, the ambient signal with average transmit power $\sigma_s^2$ can be expressed by 
\begin{eqnarray}
{\mathbf{s}}_k=[s_{1,k},\cdots,s_{L,k}]^T\in \mathbb{C}^{L\times 1}, \quad
s_{l,k}\sim \mathcal{CN}(0,\sigma_s^2),
\end{eqnarray}
where Gaussian source model with sampling period $T_s$ is assumed to design ambient signal \cite{ambB-noncoherent}.
Next, the signal broadcasted from the AP is attenuated by forward channels. The discrete-time channel impulse response between AP and tag $n$ can be expressed as \cite{ofdm2} 
\begin{eqnarray}
&\hspace{-4.8cm}\mathbf{f}_{n}=[f_{n}(1),\cdots,f_{n}(L_n^{+})]^T, \\ &\text{where } f_n (l') \sim \mathcal{CN}(0,\sigma_f^2/L_n^+), \quad
\sigma_f^2 = \frac{G_s A_e}{4\pi d^2}, \label{eqn:friis}
\end{eqnarray}
where $L_n^+$ denotes the number of multipath for tag $n \in \{1,2,\cdots,N\}$. In \eqref{eqn:friis}, $G_s$, $A_e$ and $d$, respectively, denote antenna gain at AP, effective aperture of receive antenna and tag distance from AP \cite{hybridB}.  Moreover, we assume independent and identically distributed (i.i.d.) channels among tags \cite{dyadic1, dyadic2, dyadic4, dyadic5}.  It is noteworthy that the forward channel with wide-band ambient signals (e.g., 20 MHz in Wi-Fi) generally experiences frequency-selective fading in sharp contrast to frequency-flat fading in RFID and bistatic backscatter, causing delay spreads of received signals in the forward channel $\mathbf{f}_{n}$ \cite{backfi}. To capture this characteristic, we can use Toeplitz matrix \cite{Toeplitz}, matrix-vector representations of circular-convolution operations. We define Toeplitz lower- and upper-triangular matrices as follows \cite{ofdm2}
\begin{eqnarray}
\mathbf{F}^{+}_n&\hspace{-0.1in}=\hspace{-0.1in}&\sum_{l'=1}^{L_n^+} f_n(l') (\mathbf{J}^{+}_L)^{l'-1} \in \mathbb{C}^{L\times L}, \\
\mathbf{F}^{-}_n&\hspace{-0.1in}=\hspace{-0.1in}&\sum_{l'=1}^{L_n^+} f_n(l') (\mathbf{J}^{-}_L)^{L-l'+1} \in \mathbb{C}^{L\times L}, \label{eqn:forward}
\end{eqnarray}
where the matrices $(\mathbf{J}^{+}_L)\in \mathbb{R}^{L \times L}$ and $(\mathbf{J}^{-}_L)\in \mathbb{R}^{L \times L}$, respectively, denote the Toeplitz forward shift and backward shift matrices of size $L \times L$ \cite{ofdm2, Toeplitz}. Due to delay spread, the received signal at tag $n$ at time slot $k$ can be expressed by superposition of $k$-th signals and $(k-1)$-th signals such that
\begin{eqnarray}
\mathbf{y}_{k,n}= \underbrace{\mathbf{F}^{+}_n \mathbf{s}_k}_{k\text{-th time}}+\underbrace{\mathbf{F}^{-}_n \mathbf{s}_{k-1}}_{k-1\text{-th time}} \in \mathbb{C}^{L\times 1}, \label{eqn:ISI}
\end{eqnarray}
where the second term in \eqref{eqn:ISI} constitutes ISI for the ambient signal $\mathbf{s}_k$.
Since tag relies on RF energy harvesting, it operates in active or idle state depending on the incident power \cite{ABCN-kaibin} of the received signal in \eqref{eqn:ISI}, which is expressed as
\begin{eqnarray}
E_n=\mathbb{E} \left [\sum_{k=1}^{K} \left \| \mathbf{y}_{k,n} \right \|^2 \right ]. \label{eqn:incident}
\end{eqnarray}

\begin{figure}
\centering
\includegraphics[width=8.9 cm]{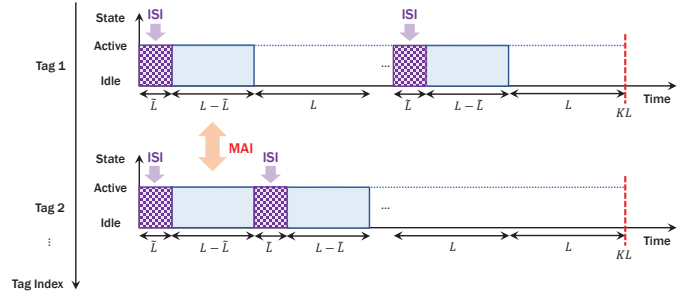}
\caption{Example of duty-cycling operation per codeword of length $\widetilde{K}=KL$.}
\label{fig:duty}
\end{figure}  

As described in Fig.~\ref{fig:duty}, the incident power in \eqref{eqn:incident} affects duty-cycling operation and thus sparsity of NOMA. Let $0<\alpha\leq 1$ be a power ratio between backscattered and incident power, affecting performance of backscatter detection as well as energy harvesting \cite{16qam,qam,codedqam}.
Let $K_1<K$ be the number of time slots when tag is activated. In a codeword of length $\widetilde{K}=KL$, each tag is activated for $K_1 L$ samples and remains idle for $(K-K_1)L$. It is remarkable that the delay spread of backscatter channel incurs ISI for $\widetilde{L} \leq L$ samples. Moreover, backscatter symbols from $N \geq K$  tags (i.e., NOMA) in active states can be interfered in time-domain causing MAI, further complicating reliable detection of AmBC. 
Since RF tags are powered by ambient energy harvesting, circuit-power constraint that harvested power with efficiency $\eta$ \cite{EH-eff} is greater than circuit power for entire $\widetilde{K}$ sampling period should be satisfied. If we assume a linear power consumption model with respect to energy per symbol $\sigma_c^2$ and backscatter symbol rate $R_s$ \cite{qam, 16qam}, the circuit-power constraint can be formulated as
\begin{eqnarray}
\underbrace{D(1-\sigma_b^2)\eta E_n}_{\text{active state}}+\underbrace{(1-D)\eta E_n}_{\text{idle state}} \geq P_c,
\end{eqnarray}
where $D = \frac{K_1}{K}$, $P_c = R_s \sigma_c^2$ and $R_s = \frac{1}{L T_s}$. The average backscatter symbol power proportional to $\alpha$ is denoted by $\sigma_b^2$ and will be discussed in Section \ref{sec:enc}. For convenient analysis of EH, we define the normalized energy per backscatter symbol as 
\begin{eqnarray}
\beta = D\sigma_b^2. \label{eqn:NEPS}
\end{eqnarray}
Under these settings, energy harvesting (EH) constraint or the circuit-power constraint can be reformulated as \cite{ABCN-kaibin}
\begin{eqnarray}
E_n \geq \theta = \frac{P_c}{\eta (1-\beta)}. \label{eqn:EH-constraint}
\end{eqnarray}
From \eqref{eqn:EH-constraint}, the state of tag $\mathbf{a}=[a_1,\cdots,a_N]$ (0: idle, 1: active) with EH threshold $\theta$ can be obtained by
\begin{eqnarray}
a_n &\hspace{-0.1in}=\hspace{-0.1in}& \mathbb{I}\{E_n \geq \theta\} , \quad \text{for } n \in \{1, \cdots, N\},
\end{eqnarray}
where $\mathbb{I} \{x \}$ denotes the indicator function of the event $x$. 
If the EH constraint is satisfied, tag can encode sparse codewords using simple modulator circuit. In the following sections, we describe AmBC modulator in Section \ref{sec:enc} and detector in Section \ref{sec:dec}.

\section{Sparse-Coded Load Modulation} \label{sec:enc}
\begin{figure}
\centering
\includegraphics[width=8.9 cm]{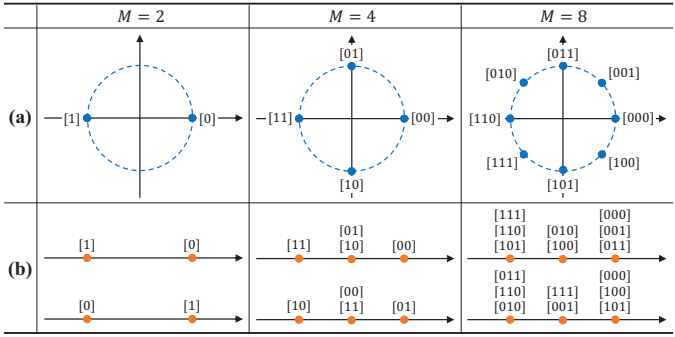}
\caption{Example of mapping function $\psi_M$ for AmBC. (a) PSK-based backscatter modulator (b) Sparse-coded backscatter modulator}
\label{fig:signal}
\end{figure}  

As described in the previous section, we design AmBC modulator based on load impedances satisfying the following characteristics:
\begin{itemize}
\item {\bf Small-form Factor}: The size of the RF tags is very limited for IoT applications such as implanted biosensors and wearables, so they should have small number of load impedances to reduce their size. The proposed SC-BMA can enable $M$-ary modulation \cite{iter-Mary,code-project} for these small-form factor IoT devices, which is in sharp contrast to traditional $M$-ary backscatter modulators with massive number of load impedances (e.g., $M$-PSK \cite{backfi}, $M$-QAM \cite{16qam, mqam1, mqam2}).
\item {\bf Interference Exploitation}: Detection performance of AmBC is severely degraded by the interferences (i.e., MAI and ISI). To rather exploit these interferences for massive connectivity, we design the codeword structure capturing sparsity \cite{scma3} of ambient EH and characteristics of DBC \cite{dyadic1, ofdm2} where two-way fading and severe ISI occur.  
\end{itemize}

Towad these goals, we describe the modulation and encoding models \cite{scma1,scma2,scma3,scma4}, which include design of mapping function \cite{iter-Mary,code-project} and sparse factor graph \cite{scma3}.

If tag $n$ is activated, the tag generates $\log_2 M$-length bit stream $\mathbf{d}_n^{\dagger}$ as follows
\begin{eqnarray}
\hspace{-0.6cm}\mathbf{d}_n^{\dagger} &\hspace{-0.1in}
\in\hspace{-0.1in}& \{\mathbf{d}^{\star}(1),\cdots, \mathbf{d}^{\star}(M) \}, \nonumber
\\ \hspace{-0.6cm}\mathbf{d}^{\star}(m)&\hspace{-0.1in}=\hspace{-0.1in}&[d^{\star}_1(m) \cdots  d^{\star}_{\log_2 M}(m)] \in \mathbb{B}^{1 \times \log_2 M}, \quad \text{if } a_n = 1.
\end{eqnarray}
To modulate the binary vector $\mathbf{d}_n^{\dagger}$ into $M$-ary backscatter symbol vector $\mathbf{b}_n$, we define the mapping function $\psi_M$ for $K_1=2$ and $M \in \{2,4,8\}$ as
\begin{eqnarray}
&\begin{matrix}
	\psi_2 = 
	\begin{cases}
	\sqrt{\alpha}[1, -1], \quad &\hspace{0.01in}\text{for } \mathbf{d}_n^{\dagger}=[0], \\
	\sqrt{\alpha}[-1,  1], \quad &\hspace{0.01in}\text{for } \mathbf{d}_n^{\dagger}=[1],
	\end{cases} \\
	\hspace{0.07in}\psi_4 =  
	\begin{cases}
	\sqrt{\alpha}[1, 0],  \quad &\text{for } \mathbf{d}_n^{\dagger}=[00], \\
	\sqrt{\alpha}[0, 1],  \quad &\text{for } \mathbf{d}_n^{\dagger}=[01], \\
	\sqrt{\alpha}[0, -1],  \quad &\text{for } \mathbf{d}_n^{\dagger}=[10], \\
	\sqrt{\alpha}[-1, 0],  \quad &\text{for } \mathbf{d}_n^{\dagger}=[11],
	\end{cases} 
	\end{matrix}
\nonumber\\ &\hspace{0.6cm}\psi_8 = 
\begin{cases}
\sqrt{\alpha}[1, 1],  \quad &\text{for } \mathbf{d}_n^{\dagger}=[000], \\
\sqrt{\alpha}[1, 0],  \quad &\text{for } \mathbf{d}_n^{\dagger}=[001], \\
\sqrt{\alpha}[0, -1],  \quad &\text{for } \mathbf{d}_n^{\dagger}=[010], \\
\sqrt{\alpha}[1, -1],  \quad &\text{for } \mathbf{d}_n^{\dagger}=[011], \\
\sqrt{\alpha}[0, 1],  \quad &\text{for } \mathbf{d}_n^{\dagger}=[100], \\
\sqrt{\alpha}[-1, 1],  \quad &\text{for } \mathbf{d}_n^{\dagger}=[101], \\
\sqrt{\alpha}[-1, -1],  \quad &\text{for } \mathbf{d}_n^{\dagger}=[110], \\
\sqrt{\alpha}[-1, 0],  \quad &\text{for } \mathbf{d}_n^{\dagger}=[111].
\end{cases}
\label{eqn:mapping}
\end{eqnarray}
Then, the $M$-ary binary vector $\mathbf{b}_n$ can be expressed by
\begin{eqnarray}
\mathbf{b}_n &\hspace{-0.1in}=\hspace{-0.1in}& \psi_M(\mathbf{d}_n^{\dagger}) \in \{\mathbf{b}^{\star}(1), \cdots, \mathbf{b}^{\star}(M) \}, 
\\
 \mathbf{b}^{\star}(m) &\hspace{-0.1in}=\hspace{-0.1in}& [b_1^{\star}(m), \cdots, b_{K_1}^{\star}(m)]  \in \mathbb{R}^{1 \times K_1}, \nonumber\\
\sigma_b^2 &\hspace{-0.1in}=\hspace{-0.1in}& \mathbb{E}\left [ \frac{\|\mathbf{b}_n \|^2}{K_1}\right ]=
\begin{cases}
\alpha, \quad &\text{for } M = 2, \\
0.5\alpha, \quad &\text{for } M = 4,\\
0.75 \alpha, \quad &\text{for } M = 8.
\end{cases} \label{eqn:backpower}
\end{eqnarray}

As described in Fig.~\ref{fig:signal}, these mapping functions satisfy the small-form factor requirement in RF tags. For instance, in conventional phase-shift keying (PSK)-based backscatter modulator, $M$ constellation points are present in signal constellation with phase difference $2\pi/M$ in Fig.~\ref{fig:signal}. (a) \cite{backfi}.
So the modulator requires $N_Z = M$ distinct load impedances, inevitably increasing size of RF tags.
On the other hand, SC-AmBC modulator only has 3 constellation points even for higher-order modulation. Accordingly, the required number of load impedances is $N_Z=2$, since zero-constellation point is equivalent to the idle state of RF tag \cite{codingB}. Although several constellation points can be overlapped in Fig.~\ref{fig:signal}. (b), symbols can be detectable as long as they are uniquely mapped in $K_1$-dimensional signal space in \eqref{eqn:mapping}, even achieving {\em modulation diversity gain} \cite{mod-div} for reliable AmBC.
Thus, high-order modulation can be implemented to typical binary PSK (BPSK) backscatter modulator, enabling high data rate for small-form factor passive RF devices such as biosensors and wearables.

In the sequel, $K_1$-length modulated symbols $\mathbf{b}_n$ are spread into $K$-length sparse codewords using one-way factor graph (Tanner graph) $\widetilde{\mathbf{G}}$ \cite{LDS}. 
By doing so, dimension of signal is extended to $2K$-dimension \cite{scma1, scma2, scma3, scma4} with $M^N$ points in main constellation, in contrast to conventional AmBC modulators typically laid on 2-dimensional Euclidean spaces \cite{wireless} with $M$ constellation points (e.g., 4-QAM \cite{qam,codedqam}, 16-QAM \cite{16qam,mqam1}, 16-PSK \cite{backfi}, 32/64-QAM \cite{mqam2}).
As a result, the number of tags allowed to transmit is increased to $N=\binom{K}{K_1}$. The overloading factor \cite{scma1}, ratio of the number of RF tags and the time slots can be defined by
\begin{eqnarray}
\lambda = N/K \geq 1.
\end{eqnarray}
In addition, the duty cycle of tag is extended to $D=K_1/K$, which is $K_1$-fold increase compared to the time-division-based AmBC (TD-AmBC) with $D=1/K$ \cite{backfi, ambB-TD}. Although $N_1=\binom{K-1}{K_1-1}$ tags are interfering in arbitrary time slot, the MAI can be efficiently suppressed by exploiting sparse code structure. We choose $K_1=2$ to achieve maximum sparsity as well as reasonable error rate for backscatter detection \cite{scma3}. 

\begin{figure}
\centering
\includegraphics[width=8.9 cm]{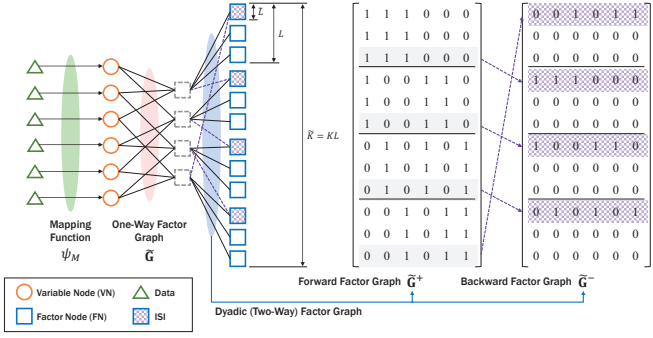}
\caption{Example of dyadic factor graph composed of forward factor graph $\widetilde{\mathbf{G}}^{+}$ and backward factor graph $\widetilde{\mathbf{G}}^{-}$ ($K=4$, $N=6$, $K_1=2$, $L=3$, $\widetilde{L}=1$).}
\label{fig:Ex_FG}
\end{figure}  

Note that in AmBC, to conserve energy as well as increase SNR for RF tags, $K$-length codewords are spanned on $L$ samples, generating $\widetilde{K}=KL$-length codewords. Also, ambient backscatter signal experiences DBC in contrast to the one-way Rayleigh channel in conventional sparse coding. Based on these observations, we introduce dyadic factor graph extending one-way factor graph $\widetilde{\mathbf{G}}$ as shown in Fig.~\ref{fig:Ex_FG}. The dyadic (two-way) factor graph has 2 kinds of nodes, namely, variable node (VN) representing codeword of tag $n \in \{1,\cdots,N\}$ and function node (FN) representing received signal at sample $\widetilde{k} \in \{1,\cdots,\widetilde{K}\}$. 
Due to ISI among backscatter signals, the dyadic factor graph is composed of the forward factor graph $\widetilde{\mathbf{G}}^{+} =\{\widetilde{g}^{+}_{\widetilde{k},n}\} \in \mathcal{B}^{\widetilde{K} \times N}$ and the backward factor graph $\widetilde{\mathbf{G}}^{-} =\{\widetilde{g}^{-}_{\widetilde{k},n}\}\in \mathcal{B}^{\widetilde{K} \times N}$ representing interference symbols. Here, we use the superscript $\{\cdot\}^-$ to represent the signals related to ISI and the superscript $\{\cdot\}^+$ vice versa.
The backward factor graph has $K\widetilde{L}$ non-zero rows which are circular-shifted from the forward factor graph, and sparse matrix under the condition $\widetilde{L}<L$, stating that symbols are only interfered by neighboring symbols \cite{ofdm2}.
The new graph structure can efficiently exploit the MAI and ISI for backscatter detection, satisfying the second requirement of AmBC for massive connectivity. Based on sparse factor graph structure \cite{scma3} in UL, the backscatter modulation can be generalized by the pseudo codes in Algorithm \ref{alg:code}.

\begin{algorithm} 
\caption{Sparse-coded backscatter modulation algorithm (SC-BMA)}
\begin{algorithmic}[1]
\State {\bf input } the number of time slots $K$ and the modulation order $M$
\State obtain the symbol mapping $\mathbf{b}^{\star}(m)=\psi_M(\mathbf{d}^{\star}(m)), \quad m \in \{1,\cdots,M\}$ from \eqref{eqn:mapping}
\For{$m \in \{1,\cdots,M\}$}
\State {\bf initialize} $n=1$
	\For{$n_1=1:K-1$}
		\For{$n_2=1:K-n_1$}
			\For{$k=1:K$}
				\If{$k=n_1$}
					\State $\widetilde{b}_{k,n}(m) = b_1^{\star}(m), \quad \widetilde{g}_{k,n} = 1$ \Comment{first non-zero element}	
				\ElsIf{$k=n_1+n_2$}
					\State $\widetilde{b}_{k,n}(m) = b_2^{\star}(m), \quad \widetilde{g}_{k,n} = 1$ \Comment{second non-zero element}
				\Else				
					\State $\widetilde{b}_{k,n}(m) = 0, \quad \widetilde{g}_{k,n} = 0$ \Comment{zero element}
				\EndIf
			\EndFor
			\State $n \leftarrow n+1$
		\EndFor
	\EndFor
\EndFor
\State obtain $\widetilde{\mathbf{G}} = \{\widetilde{g}_{k,n}\}$, $\mathbf{B}(m) = \{\widetilde{b}_{k,n}(m)\}$ \Comment{one-way factor graph, codebook}

\State generate $\widetilde{\mathbf{G}}^{+} = \widetilde{\mathbf{G}}\otimes \mathbf{1}_L$, $\mathbf{B}^{+}(m) = \mathbf{B}(m)\otimes \mathbf{1}_L$

\State generate $\widetilde{\mathbf{G}}^{-} =[(\mathbf{J}_K^{+}+(\mathbf{J}_K^{-})^{K-1})\widetilde{\mathbf{G}}] \otimes [\mathbf{1}_{\widetilde{L}}^T ~\mathbf{0}_{L-\widetilde{L}}^T]^T$, $\mathbf{B}^{-}(m)=[(\mathbf{J}_K^{+}+(\mathbf{J}_K^{-})^{K-1})\mathbf{B}(m)] \otimes [\mathbf{1}_{\widetilde{L}}^T ~\mathbf{0}_{L-\widetilde{L}}^T]^T$

\State {\bf output} the codebooks $\{\mathbf{B}^{+}(m), \mathbf{B}^{-}(m)\}_{m \in \{1,\cdots,M\}}$ and the factor graphs $\widetilde{\mathbf{G}}^{+}, \widetilde{\mathbf{G}}^{-}$
\end{algorithmic} \label{alg:code}
\end{algorithm}
Since channel information is not available at RF tags, they encode data using the one-way codebook $\{\mathbf{B}(m)\}_{m \in \{1,\cdots,M\}}$ without knowing ISI, and the forward and backward codebooks $\{\mathbf{B}^{+}(m), \mathbf{B}^{-}(m)\}_{m \in \{1,\cdots,M\}}$ containing information on ISI will be used in detection procedure in Section \ref{sec:dec}. So the reflection coefficients and backscattered signals at RF tags over $L$ samples can be represented by
\begin{eqnarray}
&\mathbf{x}_{k,n}=\Gamma_{k,n}\mathbf{y}_{k,n}\in \mathbb{C}^{1\times L}, \nonumber\\ &\text{where }
\Gamma_{k,n} \in 
\begin{cases}
\{0\},\quad &\text{if } a_n=0 \text{ (idle)}, \\  
\widetilde{b}_{k,n}(m), \quad &\text{if } a_n=1 \text{ (active)}.
\end{cases} \label{eqn:gamma}
\end{eqnarray}
Similar to the forward channel, backward channel and interference channel can be defined by
\begin{eqnarray}
\mathbf{g}_{n}&\hspace{-0.1in}=\hspace{-0.1in}&[g_{n}(1),\cdots,g_{n}(L_n^-)]^T, \quad
\mathbf{h}_0=[h_0(1),\cdots,h_0(L_0)]^T,  \nonumber \\
\mathbf{G}^{+}_n&\hspace{-0.1in}=\hspace{-0.1in}&\sum_{l'=1}^{L_n^-} g_n(l') (\mathbf{J}^{+}_L)^{l'-1} \in \mathbb{C}^{L\times L}, \nonumber\\
\mathbf{G}^{-}_n&\hspace{-0.1in}=\hspace{-0.1in}&\sum_{l'=1}^{L_n^-} g_n(l') (\mathbf{J}^{-}_L)^{L-l'+1} \in \mathbb{C}^{L\times L}, \nonumber \\
\mathbf{H}^{+}_0&\hspace{-0.1in}=\hspace{-0.1in}&\sum_{l'=1}^{L_0} h_0(l') (\mathbf{J}^{+}_L)^{l'-1} \in \mathbb{C}^{L\times L}, \nonumber\\
\mathbf{H}^{-}_0&\hspace{-0.1in}=\hspace{-0.1in}&\sum_{l'=1}^{L_0} h_0(l') (\mathbf{J}^{-}_L)^{L-l'+1} \in \mathbb{C}^{L\times L}, \label{eqn:backch}
\end{eqnarray}
where $g_n (l') \sim \mathcal{CN}(0,\sigma_g^2/L_n^-)$, $h_0 (l') \sim \mathcal{CN}(0,\sigma_{h_0}^2/L_0)$ and the variance of backward and leakage channels are denoted by $\sigma_g^2$ and $\sigma_{h_0}^2$, respectively. In \eqref{eqn:backch}, $L_n^-$ and $L_0$ denote the number of multipaths for backward channel and leakage channel, respectively.
Then, the signal received at AP at time slot $k$ can be expressed as
\begin{eqnarray}
\mathbf{r}_k &\hspace{-0.1in}=\hspace{-0.1in}& (\mathbf{H}_0^{+}\mathbf{s}_k + \mathbf{H}_0^{-}\mathbf{s}_{k-1})+
\sum_{n=1}^N (\mathbf{G}_n^{+}\mathbf{x}_{k,n}+\mathbf{G}_n^{-}\mathbf{x}_{k-1,n})+\mathbf{w}_k  \nonumber \\
&\hspace{-0.1in}=\hspace{-0.1in}&\sum_{n=1}^N \underbrace{(\mathbf{G}_n^{+}\mathbf{F}_n^{+}\mathbf{s}_{k}+
\mathbf{G}_n^{+}\mathbf{F}_n^{-}\mathbf{s}_{k-1})}_{=~\widetilde{\mathbf{h}}^{+}_{k,n}}\Gamma_{k,n} 
\nonumber\\
\quad &\hspace{-0.1in}{}\hspace{-0.1in}&+\sum_{n=1}^N \underbrace{(\mathbf{G}_n^{-}\mathbf{F}_n^{+}\mathbf{s}_{k-1}+\mathbf{G}_n^{-}\mathbf{F}_n^{-}\mathbf{s}_{k-2})}_{=~\widetilde{\mathbf{h}}^{-}_{k,n}}\Gamma_{k-1,n} 
\nonumber \\
\quad &\hspace{-0.1in}{}\hspace{-0.1in}&+ \underbrace{(\mathbf{H}_0^{+}\mathbf{s}_k+ \mathbf{H}_0^{-}\mathbf{s}_{k-1})}_{=~\widetilde{\mathbf{h}}_{k,0}} 
+\mathbf{w}_k \nonumber \\
&\hspace{-0.1in}=\hspace{-0.1in}& \underbrace{\sum_{n=1}^N \widetilde{\mathbf{h}}^{+}_{k,n}\Gamma_{k,n}}_{\text{forward codeword (MAI)}}+
\underbrace{\sum_{n=1}^N \widetilde{\mathbf{h}}^{-}_{k,n}\Gamma_{k-1,n}}_{\text{backward codeword (ISI)}}\nonumber \\
\quad &\hspace{-0.1in}{}\hspace{-0.1in}&+\underbrace{\widetilde{\mathbf{h}}_{k,0}}_{\text{SI}}+\underbrace{\mathbf{w}_k}_{\text{noise}}
\in \mathbb{C}^{L \times 1}, \label{eqn:superposition} 
\end{eqnarray}
where $\mathbf{w}_k = [w_{1,k},\cdots,w_{L,k}]^T$ denotes additive white Gaussian noise with variance $\sigma_n^2$. Since the AP knows the ambient signal $\mathbf{s}_k$ for $k \in \{1,\cdots, K\}$ and the leakage channel $\mathbf{h}_0$, the SI term in \eqref{eqn:superposition} can be reconstructed for SIC \cite{backfi}. The SIC ouput signal after hybrid digital and analog cancellation \cite{SIC} can be given by
\begin{eqnarray}
\mathbf{t}_k = \mathbf{r}_k-\widetilde{\mathbf{h}}_{k,0} \in \mathbb{C}^{L \times 1}. \label{eqn:SIC}
\end{eqnarray}
However, as discussed in \cite{backfi,ofdm2}, AP estimates the composite forward-backward channel $\{\mathbf{h}_n\}_{n \in \{1,\cdots,N\}}$ rather than the individual estimates of the forward channels $\{\mathbf{f}_n\}_{n \in \{1,\cdots,N\}}$ and the backward channels $\{\mathbf{g}_n\}_{n \in \{1,\cdots,N\}}$. The channels $\mathbf{h}_n$ can be efficiently obtained by compressed sensing based channel estimation techniques \cite{cs-survey}, which can be defined as
\begin{eqnarray}
\mathbf{h}_{n}&\hspace{-0.1in}=\hspace{-0.1in}&[h_{n}(1),\cdots,h_{n}(L_n^\star)]^T, \nonumber\\
\mathbf{H}^{+}_n&\hspace{-0.1in}=\hspace{-0.1in}&\sum_{l'=1}^{L_n^\star} h_n(l') (\mathbf{J}^{+}_L)^{l'-1} \in \mathbb{C}^{L\times L}, \nonumber\\
\mathbf{H}^{-}_n&\hspace{-0.1in}=\hspace{-0.1in}&\sum_{l'=1}^{L_n^\star} h_n(l') (\mathbf{J}^{-}_L)^{L-l'+1} \in \mathbb{C}^{L\times L},
\end{eqnarray}
where $L_n^\star = L_n^{+}+L_n^{-}-1$ denotes the number of multipaths for the composite channel.
Assuming i.i.d. fully-correlated DBC model with channel reciprocity appearing in in-band wireless-powered communication network \cite{wpcn}, the backward channel satisfies $\mathbf{g}_n = \mathbf{f}_n$ and has $L^{-} = L_n^{-} =L^{+}= L_n^{+}$ multipaths. Consequently, for AmBC detection, we should estimate the forward channels with length $L^{+}$ from the dyadic channels with $L^\star = 2 L^{+}-1$. Based on the circular convolution operation $\ast$, the relation between $\mathbf{f}_n$ and $\mathbf{h}_n$ can be represented by
\begin{eqnarray}
\hspace{-0.6cm} \mathbf{h}_n &\hspace{-0.1in}=\hspace{-0.1in}& \mathbf{f}_n \ast  \mathbf{f}_n, \nonumber\\
\hspace{-0.6cm} h_n(l') &\hspace{-0.1in}=\hspace{-0.1in}& \sum_{i'} f_n (i') f_n (l' - i' +1), \quad l' \in \{1, \cdots, L^\star \}. \label{eqn:circ-conv}
\end{eqnarray}
For example, if $L^{+}=3$, \eqref{eqn:circ-conv} can be expressed by
\begin{eqnarray}
h_n(1) &\hspace{-0.1in}=\hspace{-0.1in}& f_n(1)f_n(1)=f_n(1)^2, \label{eqn:h1}\\
h_n(2) &\hspace{-0.1in}=\hspace{-0.1in}& f_n(1)f_n(2)+f_n(2)f_n(1) = 2f_n(1)f_n(2), \label{eqn:h2} \\
h_n(3) &\hspace{-0.1in}=\hspace{-0.1in}& f_n(1)f_n(3)+f_n(2)f_n(2)+f_n(3)f_n(1)  \nonumber \\&\hspace{-0.1in}=\hspace{-0.1in}&f_n(2)^2+2f_n(1)f_n(3), \label{eqn:h3}\\
h_n(4) &\hspace{-0.1in}=\hspace{-0.1in}& f_n(2)f_n(3)+f_n(3)f_n(2) = 2f_n(2)f_n(3), \\
h_n(5) &\hspace{-0.1in}=\hspace{-0.1in}& f_n(3)f_n(3) = f_n(3)^2.
\end{eqnarray}
From \eqref{eqn:h1}, $f_n(1)$ can be given by $f_n(1) = \pm \sqrt{h_n(1)}$. we choose the estimated channel $\hat{f}_n(1) = \sqrt{h_n(1)}$ from \eqref{eqn:h1} and then sequentially solve \eqref{eqn:h2} and \eqref{eqn:h3}. From \eqref{eqn:superposition}, by solving the first $L^{+}$ equations, the vector $\hat{\mathbf{f}}_n$ and the corresponding matrices $\hat{\mathbf{F}}_n^{+}$ and $\hat{\mathbf{F}}_n^{-}$ can be obtained from \eqref{eqn:forward}. Then, the dyadic channels in \eqref{eqn:superposition} can be reformulated as
\begin{eqnarray}
\widetilde{\mathbf{h}}^{+}_{k,n} \left(\hat{\mathbf{f}}_n\right)&\hspace{-0.1in}=\hspace{-0.1in}&\hat{\mathbf{F}}_n^{+}\hat{\mathbf{F}}_n^{+}\mathbf{s}_k +\hat{\mathbf{F}}_n^{+}\hat{\mathbf{F}}_n^{-}\mathbf{s}_{k-1}, \label{eqn:comp1}\\
\widetilde{\mathbf{h}}^{-}_{k,n}\left(\hat{\mathbf{f}}_n\right)&\hspace{-0.1in}=\hspace{-0.1in}&\hat{\mathbf{F}}_n^{-}\hat{\mathbf{F}}_n^{+}\mathbf{s}_{k-1} , \label{eqn:comp2}
\end{eqnarray}
where the second term $\hat{\mathbf{F}}_n^{-}\hat{\mathbf{F}}_n^{-}\mathbf{s}_{k-2}=\mathbf{0}_{L}$ is neglected under the condition $L^\star \leq L$ \cite{ofdm2}.
\\ {\bf Remark 1}
The dyadic channels are even such that 
\begin{eqnarray}
\widetilde{\mathbf{h}}^{+}_{k,n} \left(\hat{\mathbf{f}}_n\right)&\hspace{-0.1in}=\hspace{-0.1in}&\widetilde{\mathbf{h}}^{+}_{k,n} \left(-\hat{\mathbf{f}}_n\right), \\
\widetilde{\mathbf{h}}^{-}_{k,n} \left(\hat{\mathbf{f}}_n\right)&\hspace{-0.1in}=\hspace{-0.1in}&\widetilde{\mathbf{h}}^{-}_{k,n} \left(-\hat{\mathbf{f}}_n\right).
\end{eqnarray}

Since $\hat{\mathbf{f}}_n = \pm \mathbf{f}_n$, we may incorrectly estimate the channels when $\hat{\mathbf{f}}_n = - \mathbf{f}_n$. However, the even symmetry of the channels ensures correct estimation of the channels regardless of their directions. It is noteworthy that the previous works \cite{backfi,ofdm2} in AmBC assumed $\widetilde{\mathbf{h}}^{+}_{k,n}=\mathbf{H}_n^{+}\mathbf{s}_k$ and $\widetilde{\mathbf{h}}^{-}_{k,n}=\mathbf{0}_L$ by simply discarding $L^{+}-1$ samples affected by ISI, inevitably leading to SNR reduction by factor $1-(L^{+}-1)/L$. On the other hand, the proposed channel estimation based on the Algorithm \ref{alg:estim} enables exploitation of these interferences without SNR loss.
In the following section, we discuss iterative detection scheme for superposition signals using dyadic channels as well as codebooks.
\begin{algorithm} 
\caption{Dyadic channel estimation algorithm (D-CEA)}
\begin{algorithmic}[1]
\State {\bf input } the composite forward-backward channels $\left\{\mathbf{h}_n\right\}_{n \in \{1,\cdots,N\}}$ and the ambient source signals $\left\{\mathbf{s}_k\right\}_{k \in \{1,\cdots,K\}}$
\For {$n \in \{1, \cdots, N\}$}
	\State {\bf initialize } $f_n(1) = \sqrt{h_n(1)}$ 
	\For {$l' = 2:(L^\star+1)/2$}
		\State solve \eqref{eqn:circ-conv} for the variable $f_n(l')$
	\EndFor
\State estimate the channel vector $\hat{\mathbf{f}}_n=[\hat{f}_n(1),\cdots,\hat{f}_n((L^\star+1)/2)]^T$
\State estimate the Toeplitz matrices $\hat{\mathbf{F}}_n^{+}$ and $\hat{\mathbf{F}}_n^{-}$ from the equation \eqref{eqn:forward} 
	\For {$k \in \{1,\cdots,K\}$}
		\State estimate the dyadic forward and backward channels $\widetilde{\mathbf{h}}^{+}_{k,n}$ and $\widetilde{\mathbf{h}}^{-}_{k,n}$ from \eqref{eqn:comp1}, \eqref{eqn:comp2}
	\EndFor
\EndFor
\State {\bf output } the dyadic foward channel $\left\{\widetilde{\mathbf{h}}^{+}_{k,n}\right\}_{\forall k,n}$ and the dyadic backward  channel $\left\{\widetilde{\mathbf{h}}^{-}_{k,n}\right\}_{\forall k,n}$
\end{algorithmic} \label{alg:estim}
\end{algorithm}

\begin{figure}
\centering
\includegraphics[width=8.9 cm]{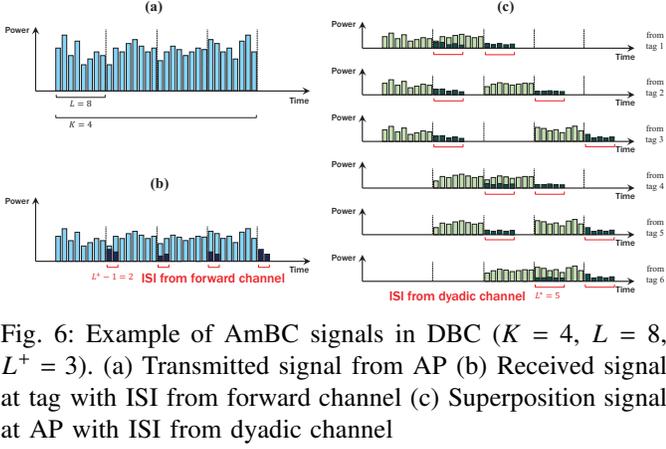}
\caption{Example of AmBC signals in DBC ($K=4$, $L=8$, $L^{+}=3$). (a) Transmitted signal from AP (b) Received signal at tag with ISI from forward channel (c) Superposition signal at AP with ISI from dyadic channel}
\label{fig:D-MPA}
\end{figure}  

\section{Iterative Detection on Dyadic Channel} \label{sec:dec}
For reliable detection of AmBC on DBC, we design the iterative detector with the following characteristics:
\begin{itemize}
\item {\bf Interference Exploitation}: To achieve massive connectivity in AmBC, the interferences (i.e., MAI and ISI) should be exploited in non-orthogonal manners \cite{scma1, FBMC}, instead of excluding them by using orthogonal signaling as in conventional detectors \cite{backfi, ofdm2}. In addition, iterative MPA should be modified for DBC to achieve additional gain without loss of backscatter signal power.
\item {\bf Low Complexity}: The received signals in AP are superposition of multiple backscatter signals with $M$-ary modulation. Also, ISI introduces another superposition of backscatter signals with delay spreads in the dyadic channels, further increasing complexity of iterative detector. Hence, the complexity of detector should be as low as possible for practical applications of SC-AmBC.
\end{itemize}

To achieve these goals, we describe iterative detector especially on DBC in this section. The SIC output \cite{SIC} in \eqref{eqn:SIC} is rewritten by superposition of $N$ reflected signals in time slot $k$ and delayed signals in time slot $k-1$, which is
\begin{eqnarray}
\mathbf{t}_k &\hspace{-0.1in}=\hspace{-0.1in}&[t_{1,k},\cdots,t_{L,k}]^T \nonumber\\
&\hspace{-0.1in}=\hspace{-0.1in}& \sum_{n=1}^{N} \widetilde{\mathbf{h}}_{k,n}^{+} \Gamma_{k,n}\nonumber\\
&\hspace{-0.1in}+\hspace{-0.1in}&\sum_{n=1}^{N} \widetilde{\mathbf{h}}_{k-1,n}^{-} \Gamma_{k-1,n}+\mathbf{w}_k\in \mathbb{C}^{L\times 1}, \\
t_{l,k} &\hspace{-0.1in}=\hspace{-0.1in}& \sum_{n=1}^{N} \widetilde{h}_{k,n,l}^{+}\Gamma_{k,n}+\sum_{n=1}^{N} \widetilde{h}_{k,n,l}^{-}\Gamma_{k-1,n}+w_{l,k} \in \mathbb{C}. \nonumber \label{eqn:SIC-SCMA}
\end{eqnarray}
Let the $n$-th codeword $\boldsymbol{\gamma}_{n}=[\Gamma_{1,n},\cdots,\Gamma_{K,n}]^T \in \mathcal{X}_n$ where $\mathcal{X}_n \subset \mathbb{C}^{K \times 1}$ is the codebook of tag $n$ with cardinality $|\mathcal{X}_n|=M$. 
Define a codeword matrix $\mathbf{X}=[\boldsymbol{\gamma}_{1},\cdots,\boldsymbol{\gamma}_{N}] \in \mathbb{C}^{K \times N}$ and a received signal matrix $\mathbf{Y}=[\mathbf{t}_1,\cdots,\mathbf{t}_K] \in \mathbb{C}^{K \times N}$. 
Then the maximum {\em a posteriori} (MAP) detection rule for codeword $\mathbf{X}$ can be expressed as \cite{scma1}
\begin{eqnarray}
\hat{\mathbf{X}} = \arg \max_{\mathbf{X}\in \mathcal{X}_1 \times\cdots \times \mathcal{X}_N} p ( \mathbf{X} | \mathbf{Y} ). \label{eqn:MAP}
\end{eqnarray}
The complexity of the above problem is shown to increases exponentially with the number of tags $N$ and polynomially with the constellation size $M$ \cite{scma1}. Fortunately, the MAP detection problem in \eqref{eqn:MAP} can be translated into marginalize product of function (MPF) problem which can be efficiently solved by MPA \cite{LDS}.
In MPA, where codewords are iteratively decoded over the factor graph, we can achieve near-optimal detection performance with low complexity \cite{scma1, scma2, scma3, scma4}. 
However, as described in Fig.~\ref{fig:D-MPA}, DBC incurs significant ISI for detection of codeword $\mathbf{X}$. If we simply filter ISI using inaccurate channel estimation as in \cite{backfi, ofdm2}, we lose large portion of signal power (e.g., maximum 62.5\% in Fig.~\ref{fig:D-MPA}. (c)) especially for high-rate AmBC applications.
Thus, to prevent the SNR loss caused by DBC, we should modify the conventional MPA \cite{LDS} to exploit the weighted-sum of ISI.
In addition, we adopt the log-domain MPA \cite{log-MAP,LDS}, which shows superior decoding performance than MPA for reliable detection of AmBC at low SNR. To achieve this goal, we define the dyadic factor graph for ambient backscatter as follows:
\begin{eqnarray}
\widetilde{\mathbf{G}}^\star =\widetilde{\mathbf{G}}^{+} \oplus \widetilde{\mathbf{G}}^{-},  
\end{eqnarray}
where the operator $\oplus$ denotes element-wise OR operation. In addition, by the channel reciprocity, $\widetilde{L} = L^{+}-1$.
\\{\bf Remark 2}
The dyadic factor graph $\widetilde{\mathbf{G}}^\star$ with $\widetilde{K}=KL$ FNs and $N$ VNs satisfies the following properties:
\begin{itemize}
\item There are $K(L-\widetilde{L})$ FNs with degree $K-1$ and $K\widetilde{L}$ FNs with degree $2K-3$.
\item There are $K$ VNs with degree $2L+\widetilde{L}$ and $N-K$ VNs with degree $2L+2\widetilde{L}$.
\end{itemize} 
If $\widetilde{L}=0$, the graph is equivalent to the spreading of the one-way factor graph $\widetilde{\mathbf{G}}$ with spreading factor $L$. As described in Fig. \ref{fig:Ex_FG}, $\widetilde{L}$ samples are collided by previous backscatter symbols causing ISI and increasing complexity of decoding. The condition $\widetilde{L} < L$ ensures low-complexity decoding since the collision is only affected by neighboring backscatter symbols \cite{ofdm2}. Especially, from Remark 2, information exchanged between FNs and VNs can be decomposed into two cases for reducing computational complexity while exploiting ISI. To perform decoding, we arrange the received signals and channels as follows
\begin{eqnarray}
\vec{\mathbf{y}}&\hspace{-0.1in}=\hspace{-0.1in}&[\vec{y}_1,\cdots,\vec{y}_{\widetilde{K}}]^T
=[\mathbf{t}_1^T, \cdots, \mathbf{t}_K^T]^T, \\
\vec{\mathbf{h}}_n^{+}&\hspace{-0.1in}=\hspace{-0.1in}&\left[\vec{h}_{1,n}^{+}, \cdots, \vec{h}_{\widetilde{K},n}^{+}\right]^T=
\left[\left(\widetilde{\mathbf{h}}^{+}_{1,n}\right)^T,\cdots,\left(\widetilde{\mathbf{h}}^{+}_{K,n}\right)^T\right]^T, \\
\vec{\mathbf{h}}_n^{-}&\hspace{-0.1in}=\hspace{-0.1in}&\left[\vec{h}_{1,n}^{-}, \cdots, \vec{h}_{\widetilde{K},n}^{-}\right]^T=
\left[\left(\widetilde{\mathbf{h}}^{-}_{1,n}\right)^T,\cdots,\left(\widetilde{\mathbf{h}}^{-}_{K,n}\right)^T\right]^T.
\end{eqnarray}
Then, we define symbol vector composed of $\widetilde{M} = \min(M,3) \leq M$ components as
\begin{eqnarray}
\vec{\mathbf{s}} = 
\begin{cases}
[\vec{S}(\widetilde{m}_1), \vec{S}(\widetilde{m}_2)] = \sqrt{\alpha}[1, -1], \quad & \text{for } M = 2,\\
[\vec{S}(\widetilde{m}_1), \vec{S}(\widetilde{m}_2), \vec{S}(\widetilde{m}_3)] = \sqrt{\alpha}[1, 0, -1], \quad & \text{for } M \geq 4.
\end{cases}
\end{eqnarray}
For notational convenience, we define the ISI condition as follows
\begin{eqnarray}
\mathcal{D}(\widetilde{k},n) = \sum_{n=1}^N a_n\widetilde{g}_{\widetilde{k},n}^{-} >0.
\end{eqnarray}
Given the received signal $\vec{\mathbf{y}}$, the channels $\{\vec{\mathbf{h}}_n^{+},\vec{\mathbf{h}}_n^{-}\}_{\forall n}$,  the noise variance $\sigma_n^2$, the tag state $\mathbf{a}$ \cite{rateless}, the factor graphs $\{\widetilde{\mathbf{G}}^{+}, \widetilde{\mathbf{G}}^{-}, \widetilde{\mathbf{G}}^{\star}\}$ and the codebooks $\{\mathbf{B}^{+}(m),\mathbf{B}^{-}(m)\}_{\forall m}$, the initial information on FN $\widetilde{k}$ can be formulated as
\begin{eqnarray}
&\hspace{-7cm}f^{\dagger}_{\widetilde{k}} (\{m_i\}_{i \in \mathcal{I}_{\widetilde{k}}})  \nonumber\\ 
&\hspace{-0.4cm}\leftarrow\begin{cases}
-\frac{1}{2\sigma_n^2} \left\|\vec{y}_{\widetilde{k}}-\sum_{i\in\mathcal{I}_{\widetilde{k}}} \left[\vec{h}_{\widetilde{k},i}^{+}B_{\widetilde{k},i}^{+}(m_{i})+ \vec{h}_{\widetilde{k},i}^{-}B_{\widetilde{k},i}^{-}(m_{i})\right]\right \|^2, \quad \\ \hspace{5cm} \text{if } \mathcal{D}(\widetilde{k},n),\label{eqn:fk_ISI}\\
-\frac{1}{2\sigma_n^2} \left\|\vec{y}_{\widetilde{k}}-\sum_{i\in\mathcal{I}_{\widetilde{k}}} \vec{h}_{\widetilde{k},i}^{+}B_{\widetilde{k},i}^{+}(m_{i})\right \|^2, \quad \\ \hspace{5cm}\text{otherwise},
\end{cases}
\label{eqn:fk}
\end{eqnarray}
where the set $\mathcal{I}_{\widetilde{k}}=\left\{n\big|a_n \widetilde{g}_{\widetilde{k},n}^{\star}>0\right\}$ denotes the VN indices connected to FN $\widetilde{k}$ and codeword indices satisfy $m_i \in \{1, \cdots, M\}$. 
If FN $\widetilde{k}$ is affected by ISI, the initial information is modified by adding the weighted-sum of ISI represented by the codebook $\{\mathbf{B}^{-}(m)\}_{\forall m}$ in \eqref{eqn:fk}. 
On the other hand, the dimension of the information in \eqref{eqn:fk} increases exponentially as modulation order $M$ increases \cite{iter-Mary, code-project}. Alternatively, we can utilize the fact that the number of symbols is much smaller than the number of codewords as described in Fig.~\ref{fig:signal}. For example, if $M=8$, total 8 codewords are represented in $K_1=2$ dimension with 3 overlapped symbols. 
The non-orthogonal projections of the mapping function enable low-complexity decoding even for higher values of $M$ by compression of the information.
Using codeword projection method \cite{iter-Mary, code-project}, the information in \eqref{eqn:fk} can be reformulated as
\begin{eqnarray}
&\hspace{-7cm}\vec{f}_{\widetilde{k}}^{-} (\{{m}_i\}_{i \in \mathcal{I}_{\widetilde{k}}}) 
\nonumber \\ 
&\leftarrow -\frac{1}{2\sigma_n^2} \left\|\vec{y}_{\widetilde{k}}-\sum_{i\in\mathcal{I}_{\widetilde{k}}} \left[\vec{h}_{\widetilde{k},i}^{+}B_{\widetilde{k},i}^{+}(m_{i})+ \vec{h}_{\widetilde{k},i}^{-}B_{\widetilde{k},i}^{-}(m_{i})\right]\right \|^2, \nonumber\\ 
&\text{if } \mathcal{D}(\widetilde{k},n), \\
&\hspace{-7cm}\vec{f}_{\widetilde{k}}^{+} (\{\widetilde{m}_i\}_{i \in \mathcal{I}_{\widetilde{k}}}) \nonumber\\
&\hspace{-2.9cm}\leftarrow -\frac{1}{2\sigma_n^2} \left\|\vec{y}_{\widetilde{k}}-\sum_{i\in\mathcal{I}_{\widetilde{k}}} \vec{h}_{\widetilde{k},i}^{+}\vec{S}(\widetilde{m}_{i})\right \|^2, \nonumber \\ 
&\text{otherwise}. \label{eqn:nonISI}
\end{eqnarray}
Now, computational complexity at each FN is reduced for the non-ISI condition in \eqref{eqn:nonISI}.
Then, we assign {\em a priori} probability $ap_{n}(m_{n})$ to initial information from VN $n$ to FN $\widetilde{k}$ as follows:
\begin{eqnarray}
\mathrm{I}_{n\rightarrow \widetilde{k}}(m_{n})\leftarrow ap_{n}(m_{n})=\frac{1}{M}.
\end{eqnarray}
In the log-domain MPA, cascades of logarithm, summation and exponential operation are approximated using Jacobian logarithm and correction function for low-SNR applications. Now, we define a function $\overset{\star}{\max}(\delta_1,\delta_2)$ as \cite{log-MAP,LDS}
\begin{eqnarray}
\ln (e^{\delta_1}+e^{\delta_2}) \approx \overset{\star}{\max}(\delta_1,\delta_2) 
&\hspace{-0.1in}=\hspace{-0.1in}&\max (\delta_1,\delta_2) \nonumber\\
&\hspace{-0.1in}+\hspace{-0.1in}&\ln (1+e^{-|\delta_2-\delta_1|}).
\end{eqnarray}
In the next step, FN $\widetilde{k}$ passes the updated messages to its neighboring VNs. The intermediate variable for the log-domain MPA can be calculated as
\begin{eqnarray}
&\hspace{-6.5cm}\widetilde{\mathrm{I}}^{-}_{\widetilde{k}\rightarrow n} (\{{m}_i\}_{i \in \mathcal{I}_{\widetilde{k}}}) \nonumber\\
&\hspace{-0.5cm}\leftarrow
\vec{f}^{-}_{\widetilde{k}} ( (\{{m}_i\}_{i \in \mathcal{I}_{\widetilde{k}}}))
+\sum_{i \in \mathcal{I}_{\widetilde{k}} \setminus \{n\}}
{\mathrm{I}}_{i\rightarrow \widetilde{k}}({m}_{i}),  \quad \text{if } \mathcal{D}(\widetilde{k},n), \\
&\hspace{-6.5cm}\widetilde{\mathrm{I}}^{+}_{\widetilde{k}\rightarrow n} (\{\widetilde{m}_i\}_{i \in \mathcal{I}_{\widetilde{k}}})\nonumber \\
&\hspace{-0.5cm}\leftarrow
\vec{f}^{+}_{\widetilde{k}} ( (\{\widetilde{m}_i\}_{i \in \mathcal{I}_{\widetilde{k}}}))
+\sum_{i \in \mathcal{I}_{\widetilde{k}} \setminus \{n\}}
\bar{\mathrm{I}}_{i\rightarrow \widetilde{k}}(\widetilde{m}_{i}), \quad \text{otherwise},
\end{eqnarray}
where $\bar{\mathrm{I}}_{n\rightarrow \widetilde{k}}(\widetilde{m}_{n})$ denotes the projected information from VN $n$ to FN $\widetilde{k}$. To adapt previous works \cite{iter-Mary, code-project} for AmBC over $L$ samples, we define FN indicator function as follows
\begin{eqnarray}
\Lambda(\widetilde{k},n)=\left\lfloor \sum_{i=1}^{\widetilde{k}} \widetilde{g}_{i,n}^{+}/L\right\rfloor+1.
\end{eqnarray}
Consequently, the information $\bar{\mathrm{I}}_{n\rightarrow \widetilde{k}}(\widetilde{m}_{n})$ is summarized in Table \ref{table:proj_ex}. 
\begin{table*}
\small
\centering
\caption{Table of Information Projections and Expansions.}
\label{table:proj_ex}
\renewcommand{\arraystretch}{1.2} \setlength{\tabcolsep}{12pt}
\begin{tabular}{ccl}
\hline
{$M$} & {$\Lambda(\widetilde{k},n)$} & {\bf Projected Information from VN $n$ to FN $\widetilde{k}$} \\
\hline
\multirow{2}{*}{4} & $1$ & $\bar{\mathrm{I}}_{n \rightarrow \widetilde{k}} (1)=\mathrm{I}_{n \rightarrow \widetilde{k}} (1), \quad \bar{\mathrm{I}}_{n \rightarrow \widetilde{k}} (2)=\overset{\star}{\max}\left (\mathrm{I}_{n \rightarrow \widetilde{k}} (2),\mathrm{I}_{n \rightarrow \widetilde{k}} (3) \right),\quad \bar{\mathrm{I}}_{n \rightarrow \widetilde{k}} (3)=\mathrm{I}_{n \rightarrow \widetilde{k}} (4)$ \smallskip\\
{} & 2 & $\bar{\mathrm{I}}_{n \rightarrow \widetilde{k}} (1)=\mathrm{I}_{n \rightarrow \widetilde{k}} (2), \quad \bar{\mathrm{I}}_{n \rightarrow \widetilde{k}} (2)=\overset{\star}{\max}\left (\mathrm{I}_{n \rightarrow \widetilde{k}} (1),\mathrm{I}_{n \rightarrow \widetilde{k}} (4) \right),\quad \bar{\mathrm{I}}_{n \rightarrow \widetilde{k}} (3)=\mathrm{I}_{n \rightarrow \widetilde{k}} (3)$ \smallskip \\
\hline
\multirow{4}{*}{8} & \multirow{2}{*}{1} & $\bar{\mathrm{I}}_{n \rightarrow \widetilde{k}} (1)=\overset{\star}{\max} \left(\mathrm{I}_{n \rightarrow \widetilde{k}} (1),\mathrm{I}_{n \rightarrow \widetilde{k}} (2),\mathrm{I}_{n \rightarrow \widetilde{k}} (4)\right), 
\quad \bar{\mathrm{I}}_{n \rightarrow \widetilde{k}} (2)=\overset{\star}{\max}\left (\mathrm{I}_{n \rightarrow \widetilde{k}} (3),\mathrm{I}_{n \rightarrow \widetilde{k}} (5) \right),$ \\
{} & {} & $\bar{\mathrm{I}}_{n \rightarrow \widetilde{k}} (3)=\overset{\star}{\max} \left(\mathrm{I}_{n \rightarrow \widetilde{k}} (6),\mathrm{I}_{n \rightarrow \widetilde{k}} (7),\mathrm{I}_{n \rightarrow \widetilde{k}} (8)\right)$ \smallskip\\
{} & \multirow{2}{*}{2} & $\bar{\mathrm{I}}_{n \rightarrow \widetilde{k}} (1)=\overset{\star}{\max} \left(\mathrm{I}_{n \rightarrow \widetilde{k}} (1),\mathrm{I}_{n \rightarrow \widetilde{k}} (5),\mathrm{I}_{n \rightarrow \widetilde{k}} (6)\right), 
\quad \bar{\mathrm{I}}_{n \rightarrow \widetilde{k}} (2)=\overset{\star}{\max}\left (\mathrm{I}_{n \rightarrow \widetilde{k}} (2),\mathrm{I}_{n \rightarrow \widetilde{k}} (8) \right),$ \\
{} & {} & $\bar{\mathrm{I}}_{n \rightarrow \widetilde{k}} (3)=\overset{\star}{\max} \left(\mathrm{I}_{n \rightarrow \widetilde{k}} (3),\mathrm{I}_{n \rightarrow \widetilde{k}} (4),\mathrm{I}_{n \rightarrow \widetilde{k}} (7)\right)$ \smallskip \\
\hline
{$M$} & {$\Lambda(\widetilde{k},n)$} & {\bf Expanded Information from FN $\widetilde{k}$ to VN $n$}\\
\hline
\multirow{2}{*}{4} & 1 & $\bar{\mathrm{I}}_{\widetilde{k} \rightarrow n} (1)=\vec{\mathrm{I}}^{+}_{\widetilde{k} \rightarrow n}(1), \quad \bar{\mathrm{I}}_{\widetilde{k} \rightarrow n} (2)=\vec{\mathrm{I}}^{+}_{\widetilde{k} \rightarrow n}(2), \quad \bar{\mathrm{I}}_{\widetilde{k} \rightarrow n} (3)=\vec{\mathrm{I}}^{+}_{\widetilde{k} \rightarrow n}(2), \quad \bar{\mathrm{I}}_{\widetilde{k} \rightarrow n} (4)=\vec{\mathrm{I}}^{+}_{\widetilde{k} \rightarrow n}(3)$ \smallskip\\
{} & 2 & $\bar{\mathrm{I}}_{\widetilde{k} \rightarrow n} (1)=\vec{\mathrm{I}}^{+}_{\widetilde{k} \rightarrow n}(2), \quad \bar{\mathrm{I}}_{\widetilde{k} \rightarrow n} (2)=\vec{\mathrm{I}}^{+}_{\widetilde{k} \rightarrow n}(1), \quad \bar{\mathrm{I}}_{\widetilde{k} \rightarrow n} (3)=\vec{\mathrm{I}}^{+}_{\widetilde{k} \rightarrow n}(3), \quad \bar{\mathrm{I}}_{\widetilde{k} \rightarrow n} (4)=\vec{\mathrm{I}}^{+}_{\widetilde{k} \rightarrow n}(2)$ \smallskip\\
\hline
\multirow{4}{*}{8} & \multirow{2}{*}{1} & $\bar{\mathrm{I}}_{\widetilde{k} \rightarrow n} (1)=\vec{\mathrm{I}}^{+}_{\widetilde{k} \rightarrow n}(1), \quad \bar{\mathrm{I}}_{\widetilde{k} \rightarrow n} (2)=\vec{\mathrm{I}}^{+}_{\widetilde{k} \rightarrow n}(1), \quad \bar{\mathrm{I}}_{\widetilde{k} \rightarrow n} (3)=\vec{\mathrm{I}}^{+}_{\widetilde{k} \rightarrow n}(2), \quad \bar{\mathrm{I}}_{\widetilde{k} \rightarrow n} (4)=\vec{\mathrm{I}}^{+}_{\widetilde{k} \rightarrow n}(1),$ \\
{} & {} & $\bar{\mathrm{I}}_{\widetilde{k} \rightarrow n} (5)=\vec{\mathrm{I}}^{+}_{\widetilde{k} \rightarrow n}(2), \quad \bar{\mathrm{I}}_{\widetilde{k} \rightarrow n} (6)=\vec{\mathrm{I}}^{+}_{\widetilde{k} \rightarrow n}(3), \quad \bar{\mathrm{I}}_{\widetilde{k} \rightarrow n} (7)=\vec{\mathrm{I}}^{+}_{\widetilde{k} \rightarrow n}(3), \quad \bar{\mathrm{I}}_{\widetilde{k} \rightarrow n} (8)=\vec{\mathrm{I}}^{+}_{\widetilde{k} \rightarrow n}(3)$ \smallskip \\
{} & \multirow{2}{*}{2} & $\bar{\mathrm{I}}_{\widetilde{k} \rightarrow n} (1)=\vec{\mathrm{I}}^{+}_{\widetilde{k} \rightarrow n}(1), \quad \bar{\mathrm{I}}_{\widetilde{k} \rightarrow n} (2)=\vec{\mathrm{I}}^{+}_{\widetilde{k} \rightarrow n}(2), \quad \bar{\mathrm{I}}_{\widetilde{k} \rightarrow n} (3)=\vec{\mathrm{I}}^{+}_{\widetilde{k} \rightarrow n}(3), \quad \bar{\mathrm{I}}_{\widetilde{k} \rightarrow n} (4)=\vec{\mathrm{I}}^{+}_{\widetilde{k} \rightarrow n}(3),$ \\
{} & {} & $\bar{\mathrm{I}}_{\widetilde{k} \rightarrow n} (5)=\vec{\mathrm{I}}^{+}_{\widetilde{k} \rightarrow n}(1), \quad \bar{\mathrm{I}}_{\widetilde{k} \rightarrow n} (6)=\vec{\mathrm{I}}^{+}_{\widetilde{k} \rightarrow n}(1), \quad \bar{\mathrm{I}}_{\widetilde{k} \rightarrow n} (7)=\vec{\mathrm{I}}^{+}_{\widetilde{k} \rightarrow n}(3), \quad \bar{\mathrm{I}}_{\widetilde{k} \rightarrow n} (8)=\vec{\mathrm{I}}^{+}_{\widetilde{k} \rightarrow n}(2)$\smallskip\\
\hline
\end{tabular}
\end{table*}
Then, the updated message sent to neighboring VNs from FN $\widetilde{k}$ can be efficiently approximated as
\begin{eqnarray}
&\hspace{-7.2cm}\vec{\mathrm{I}}_{\widetilde{k}\rightarrow {n}}^{-}(m_{n}) \nonumber\\ &\leftarrow
\overset{\star}{\max}_{m_{i},i \in \mathcal{I}_{\widetilde{k}} \setminus \{n\}}\left(\widetilde{\mathrm{I}}^{-}_{\widetilde{k}\rightarrow n} (\{{m}_i\}_{i \in \mathcal{I}_{\widetilde{k}}})\right),  \quad \text{if } \mathcal{D}(\widetilde{k},n),  \label{eqn:Ifvm}\\
&\hspace{-7.2cm}\vec{\mathrm{I}}_{\widetilde{k}\rightarrow {n}}^{+}(\widetilde{m}_{n}) \nonumber \\
&\leftarrow
\overset{\star}{\max}_{\widetilde{m}_{i},i \in \mathcal{I}_{\widetilde{k}} \setminus \{n\}}\left(\widetilde{\mathrm{I}}^{+}_{\widetilde{k}\rightarrow n} (\{\widetilde{m}_i\}_{i \in \mathcal{I}_{\widetilde{k}}})\right),  \quad \text{otherwise}. 
 \label{eqn:Ifvp}
\end{eqnarray}
Based on the information \eqref{eqn:Ifvm} and \eqref{eqn:Ifvp}, VN $n$ sends the updated message to FN $\widetilde{k}$. Assuming $K_1=2$ and denoting the indices of FNs connected to VN $n$ as $\mathcal{J}_n$, the updated message can be represented by
\begin{eqnarray}
\mathrm{I}_{{n}\rightarrow \widetilde{k}}(m_n)&\hspace{-0.1in}\leftarrow \hspace{-0.1in}&
\ln ap_{n}(m_n)
+\mathrm{I}_{\widetilde{k}\rightarrow {n}}(m_{n})
\nonumber \\
&\hspace{-0.1in}-\hspace{-0.1in}&\overset{\star}{\max}_{m_n, i \in \mathcal{J}_n \setminus \{\widetilde{k}\}}\left(\mathrm{I}_{i\rightarrow {n}}(m_{n})\right)
\label{eqn:Ivf},
\end{eqnarray}
where the information from FN $\widetilde{k}$ to VN $n$ can be expressed by
\begin{eqnarray}
\mathrm{I}_{\widetilde{k}\rightarrow {n}}(m_n)=
\begin{cases}
\vec{\mathrm{I}}_{\widetilde{k}\rightarrow {n}}^{-}(m_{n}), \quad \text{if } \mathcal{D}(\widetilde{k},n), \\
\bar{\mathrm{I}}_{\widetilde{k}\rightarrow {n}}(m_{n}), \quad \text{otherwise},
\end{cases}
\end{eqnarray}
where $\bar{\mathrm{I}}_{\widetilde{k}\rightarrow {n}}(m_{n})$ denotes the expanded information which is also summarized in Table \ref{table:proj_ex}.

After $N_I$ iterations, VN $n$ estimates logarithm of the probability of the code $m_n$ as 
\begin{eqnarray}
Q_n (m_n) \leftarrow \ln ap_{n}(m_n) + \sum_{i \in \mathcal{J}_n} \mathrm{I}_{i\rightarrow {n}}(m_{n}).
\end{eqnarray}
Finally, the log-likelihood ratio (LLR) for $m^{\dagger}$-th most significant bit (MSB) where $m^{\dagger} \in \{1,\cdots,\log_2 M\}$ for tag $n$ can be calculated as \cite{LDS}
\begin{eqnarray}
\Theta_{n,m^{\dagger}} &\hspace{-0.1in}\leftarrow\hspace{-0.1in}& \ln \left( \frac{p(d^{\dagger}_{n,m^{\dagger}})=0}{p(d^{\dagger}_{n,m^{\dagger}})=1}\right)
\nonumber \\&\hspace{-0.1in}=\hspace{-0.1in}& \overset{\star}{\max}_{m':d^{\star}_{m^{\dagger}}(m')=0}\left( Q_n (m')\right)
\nonumber \\&\hspace{-0.1in}-\hspace{-0.1in}&\overset{\star}{\max}_{m':d^{\star}_{m^{\dagger}}(m')=1}\left( Q_n (m')\right).
\end{eqnarray}
Then, backscatter data can be detected as
\begin{eqnarray}
\hat{d}_{n,m^{\dagger}}^{\dagger}=
\begin{cases}
1,\quad &\text{if }  \Theta_{n,m^{\dagger}} \leq 0\\
0,\quad &\text{if }  \Theta_{n,m^{\dagger}} > 0
\end{cases}, \quad \text{for } a_n = 1.
\end{eqnarray}
Based on the iterative detector utilizing MAI and ISI, high-rate backscatter can be achievable by exploiting inherent characteristic of sparsity in ambient energy harvesting as well as DBC providing additional diversity gain.
\\{\bf Lemma 1}
The worst-case complexity of the iterative detector is given by
\begin{eqnarray}
\mathcal{O}\left(\left(1-\frac{\widetilde{L}}{L}\right){\widetilde{M}}^{K-1}+\frac{\widetilde{L}}{L}M^{2K-3}\right) \label{eqn:complexity}
\end{eqnarray}
\begin{IEEEproof}
Information from FN to VN can be decomposed into two cases. In the first case where ISI is not affected, $K(L-\widetilde{L})$ FNs are connected to $K-1$ VNs, respectively, and $M$-ary codeword information is passed to VNs, resulting $M^{K-1}$ combination of information. If $M$-ary codeword is projected to $\widetilde{M}$ symbols, the size of information can be reduced to $\widetilde{M}^{K-1}$ using information projection \cite{iter-Mary, code-project}. On the other hand, $K\widetilde{L}$ FNs affected by ISI are connected to additional $K-2$ VNs in dyadic factor graph, resulting $M^{2K-3}$ combination of information. Since RF tag relies on energy harvesting, the complexity of detector diminishes as density of active tags becomes sparser.
\end{IEEEproof}
As expressed in \eqref{eqn:complexity}, ISI increases complexity of the iterative detector for achieving additional diversity gain. However, the fraction of the FNs affected by the interferences is very small when $\widetilde{L} << L$ for sufficiently long period of backscatter symbols. Therefore, we can enjoy low-complexity detection even for non-orthogonal and high-order AmBC.
At last, the average sum throughput of the proposed SC-AmBC can be evaluated as
\begin{eqnarray}
C &\hspace{-0.1in}=\hspace{-0.1in}&\lambda R_b p \left (\hat{ \mathcal{I}} \big| \mathcal{E}\right ) p \left ( \mathcal{E}  \right),
\end{eqnarray}
where $R_b=R_s/K\log_2 M$ denotes the bit rate of RF tags and the successful information decoding event and energy harvesting event are denoted by $\hat{\mathcal{I}}$ and $\mathcal{E}$, respectively. The probability of successful energy harvesting and information decoding can be calculated as
\begin{eqnarray}
p\left(\hat{\mathcal{I}} \big | \mathcal{E} \right) &\hspace{-0.1in}=\hspace{-0.1in}& \mathbb{E}_{a_n=1} \left [
\frac{\left\|\hat{\mathbf{d}}_{n}^{\dagger} \odot \mathbf{d}_n^{\dagger} \right \|^2}{\log_2 M}\right], \label{eqn:BER}\\
p\left(\mathcal{E} \right) &\hspace{-0.1in}=\hspace{-0.1in}& \mathbb{E} \left [a_n\right], \label{eqn:HP}
\end{eqnarray}
where $\odot$ denotes exclusive-NOR operator. 
In conclusion, the comparison of the conventional TD-AmBC and the proposed SC-AmBC is summarized in Table~\ref{table:comparison}.

\begin{table*}
\small 
\centering
\caption{Comparison of the conventional TD-AmBC and the proposed SC-AmBC.}
\renewcommand{\arraystretch}{1.2} \setlength{\tabcolsep}{24pt}
\begin{tabular}{l  l  c c}
\hline
{} & {} & {\textbf{TD-AmBC}} & {\textbf{SC-AmBC}}\\
\hline
{\bf Modulation} & Load impedances & $M$ & $2$ \\
{} & Constellation dimension & $2$ & $2K$ \\ 
{} & Constellation symbols & $M$ & $\widetilde{M}$ \\
\hline
{\bf Detection} & ISI exploitation & no & yes \\
{} & MAI exploitation & no & yes \\
{} & Complexity & low & medium \\
{} & BER & high & low \\
\hline
{\bf Networking} & Max. sum throughput & $R_b$ & $\lambda R_b$ \\
{} & Max. connectivity & $K$ & $\binom{K}{K_1}$ \\
{} & Duty cycle & $1/K$ & $K_1/K$ \\
{} & Latency & high & low \\
{} & Multiple access & orthogonal & non-orthogonal \\
\hline
\end{tabular}
\label{table:comparison}
\end{table*}

\section{Simulation Results}\label{sec:simulation}
In this section, we evaluate the performance of our proposed SC-AmBC along with the conventional TD-AmBC. The simulation scenario is set as a wireless local area network (WLAN) using 2.4 GHz industrial, scientific and medical (ISM) band \cite{802.11}. Applications for the AmBC would be smart home or IoT environments where massive numbers of RF devices are seamlessly integrated into existing wireless infrastructure.
For AmBC, AP transmits IEEE 802.11g Wi-Fi OFDM signals \cite{backfi} while tags harvest these ambient RF signals \cite{EH-eff} and upload data streams by switching load impedances \cite{qam}. 
Then, RF tag in SC-AmBC modulates $\widetilde{M}$ load impedances with non-orthogonal sparse codeword, while that in TD-AmBC modulates $M$ impedances with TDMA \cite{backfi,ofdm2}.
Then, the AP in SC-AmBC receives the illuminated signals experiencing fully-correlated DBC \cite{dyadic1,dyadic2,dyadic3,dyadic4,dyadic5,dyadic6,ofdm2} and detects data streams from these signals via the proposed D-MPA. For simplicity, we assume a system model with single antenna at AP and no forward-error correction (FEC) codes but extensions of our model to such scenarios are straightforward manner. 
The AmBC is numerically evaluated by Monte-Carlo simulations, since it is challenging to derive a closed-form theoretical performance in multi-dimensional constellation \cite{scma3} and in the DBC \cite{dyadic6} especially for low SNR.
System parameters are summarized in Table~\ref{table:general_parameters}.

 \begin{table*}
\small
\centering
\caption{System Paramters.}
\label{table:general_parameters}
\renewcommand{\arraystretch}{1.2} \setlength{\tabcolsep}{18pt}
\begin{tabular}{l  l  l}
\hline
{\textbf{Parameters}} & {\textbf{Values}} & {\textbf{References}}\\
\hline
OFDM signaling & $\sigma_s^2 = 20~{\mathrm{dBm}} ~/~  \sigma_n^2 = -90~{\mathrm{dBm}} ~/~T_s = 50~\mathrm{ns} $ & IEEE 802.11g \cite{802.11}\\
\hline
Antenna property & $A_e = 0.0012 ~\mathrm{m}^2~/~ G_s = +3~\mathrm{dB}$ & Testbed results \cite{backfi}\\
\hline
Tag hardware & $\eta = 0.25 ~/~ \sigma_c^2 = 0.58 ~\mathrm{pJ/symbol}$ & Testbed results \cite{qam,EH-eff}\\
\hline
Sparse code & $K_1 = 2 ~/~ N_I = 3~\mathrm{iterations}$ & Theoretical results \cite{scma3} \\
\hline
Coverage, rate & $d = 5~\mathrm{m}$ ~/~$L=8~\mathrm{samples}$ & Simulation settings \cite{backfi} \\
\hline
\end{tabular}
\end{table*}

\subsection{Applicability to $M$-ary Modulation}
\begin{figure*}
\begin{center}
\parbox{7.5cm}
{\small \centering \includegraphics [width=7.5cm]{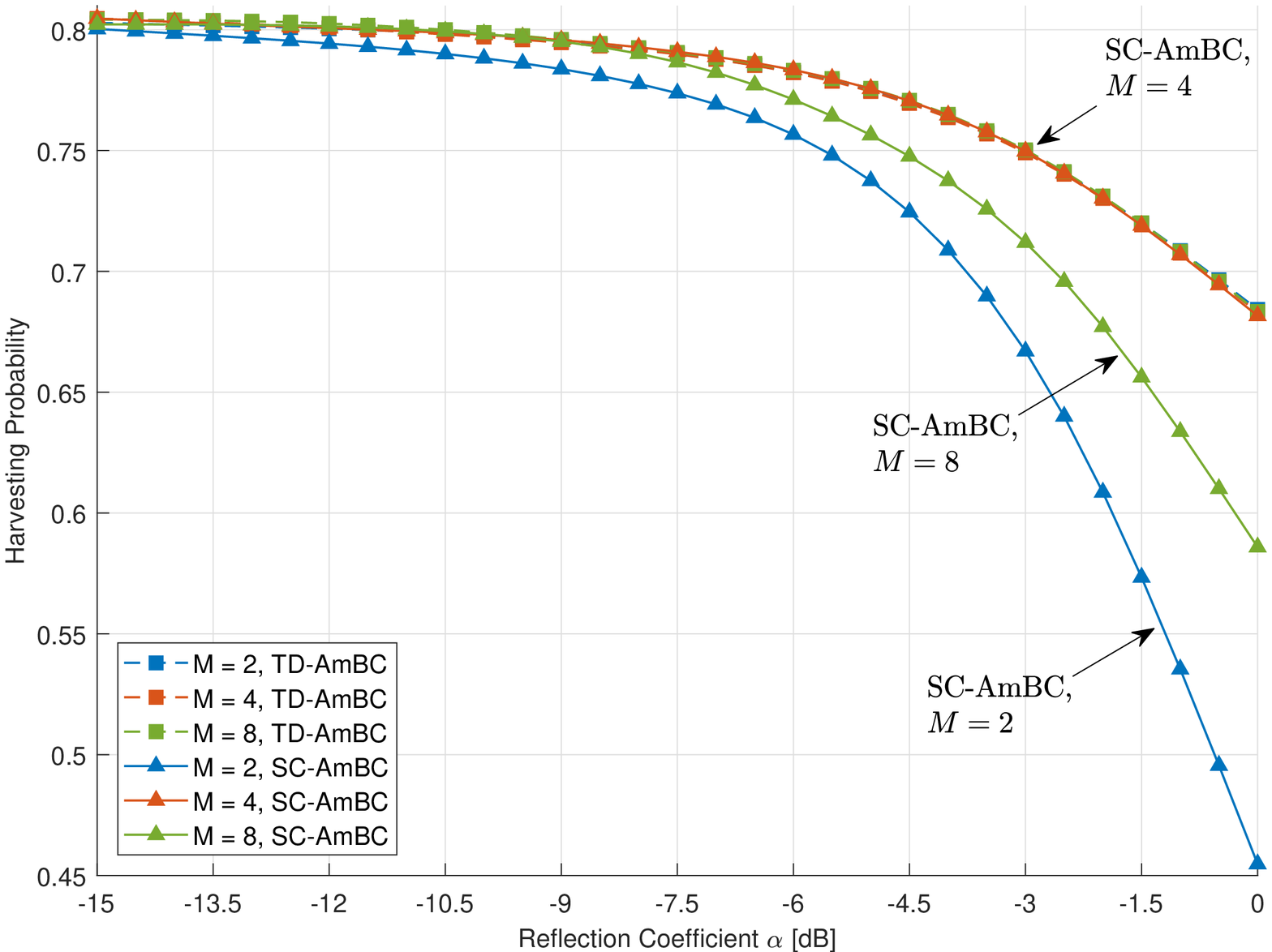} \\
(a) Harvesting probability $p\left(\mathcal{E} \right)$}
\quad
\parbox{7.5cm}
{\small \centering \includegraphics [width=7.5cm]{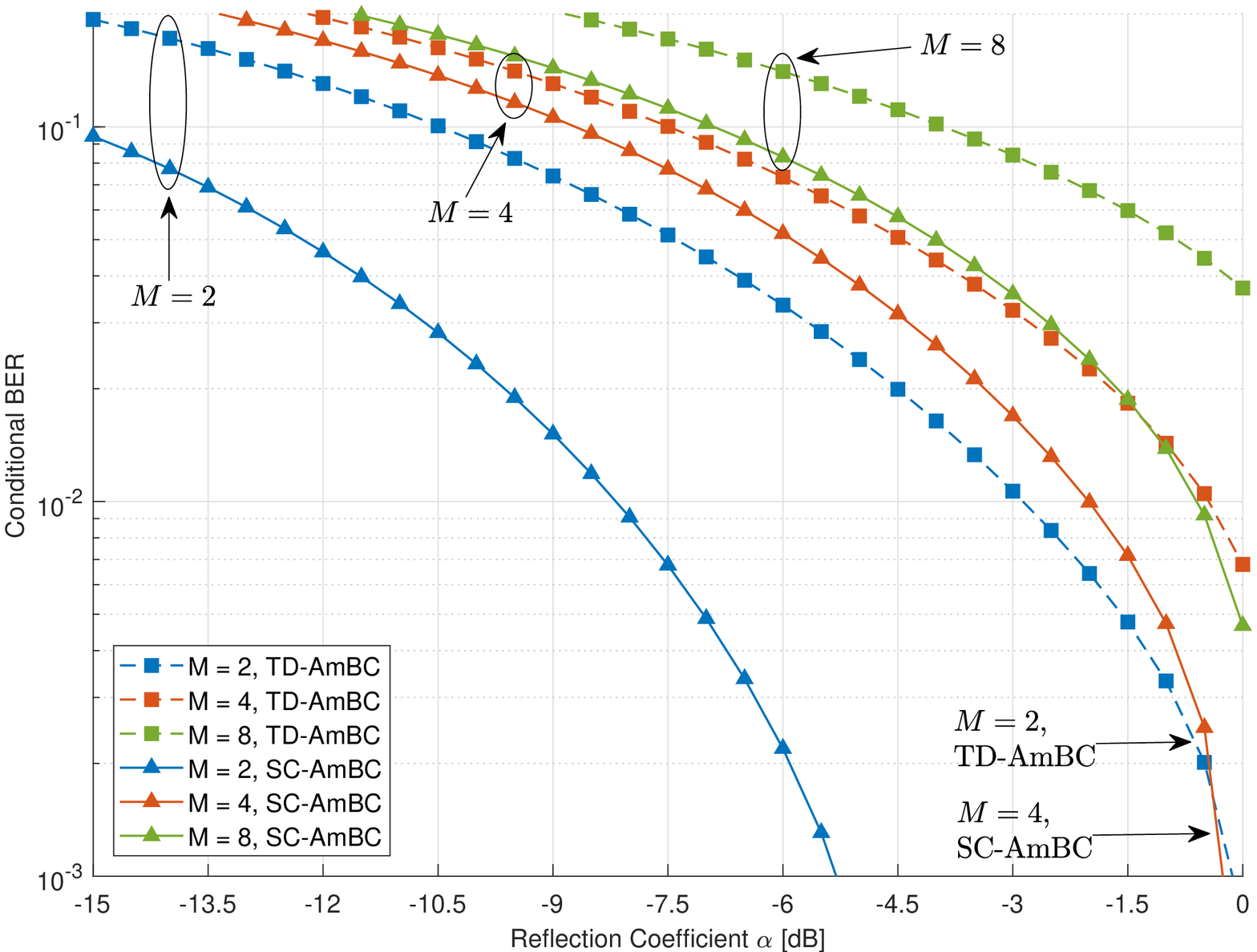}
(b) Conditional BER $1-p\left(\hat{\mathcal{I}} \big | \mathcal{E} \right)$}
\vspace{-0.0cm}
\caption{Effect of the modulation order $M$ ($L^{+} = 3$, $K = 4$, $\lambda=1.5$ for SC-AmBC).}
\label{fig:mod}
\end{center}
\end{figure*}

\begin{figure*}[h]
\begin{center}
\parbox{7.5cm}
{\small \centering \includegraphics [width=7.5cm]{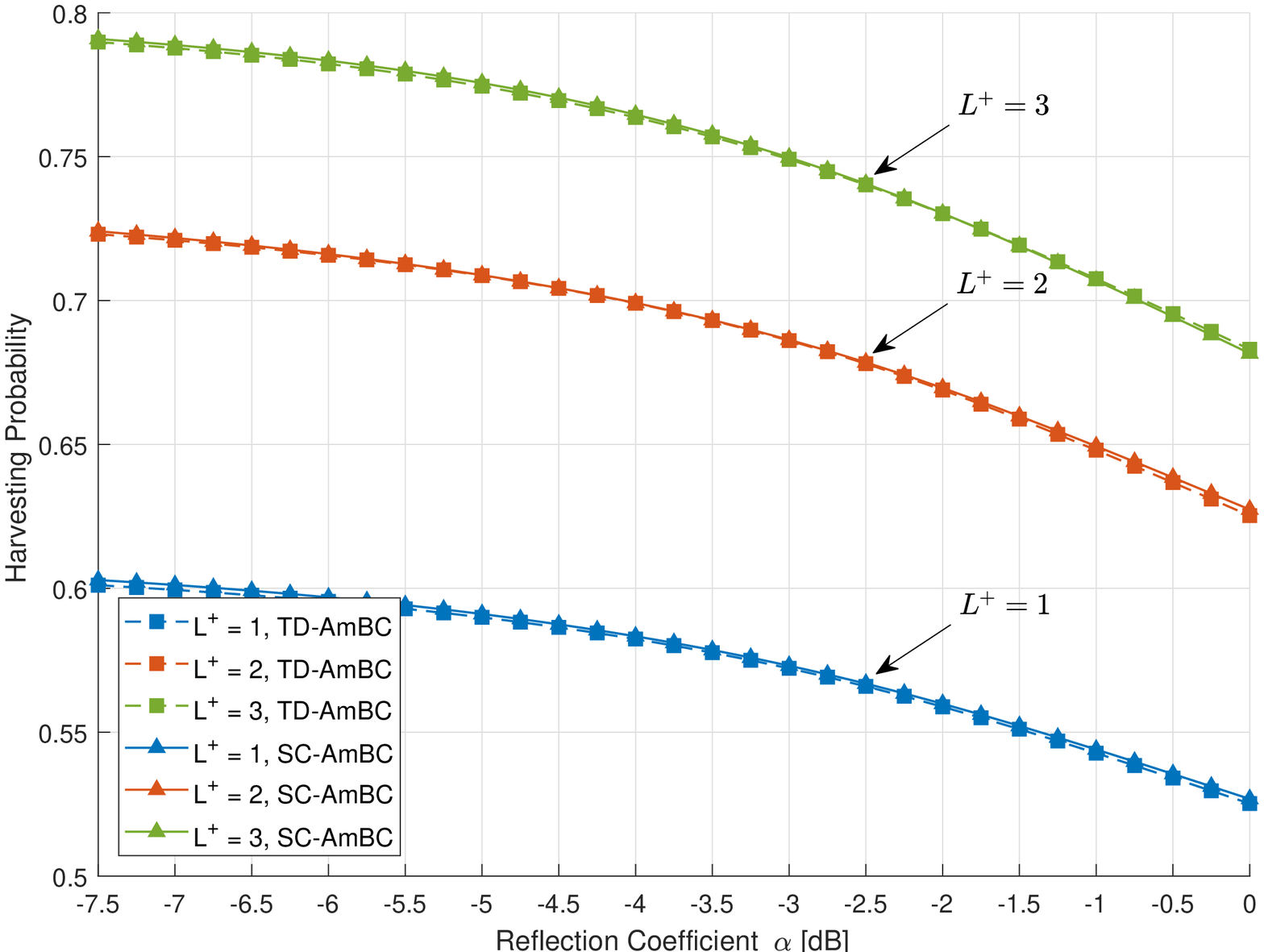} \\
(a) Harvesting probability $p\left(\mathcal{E} \right)$}
\quad
\parbox{7.5cm}
{\small \centering \includegraphics [width=7.5cm]{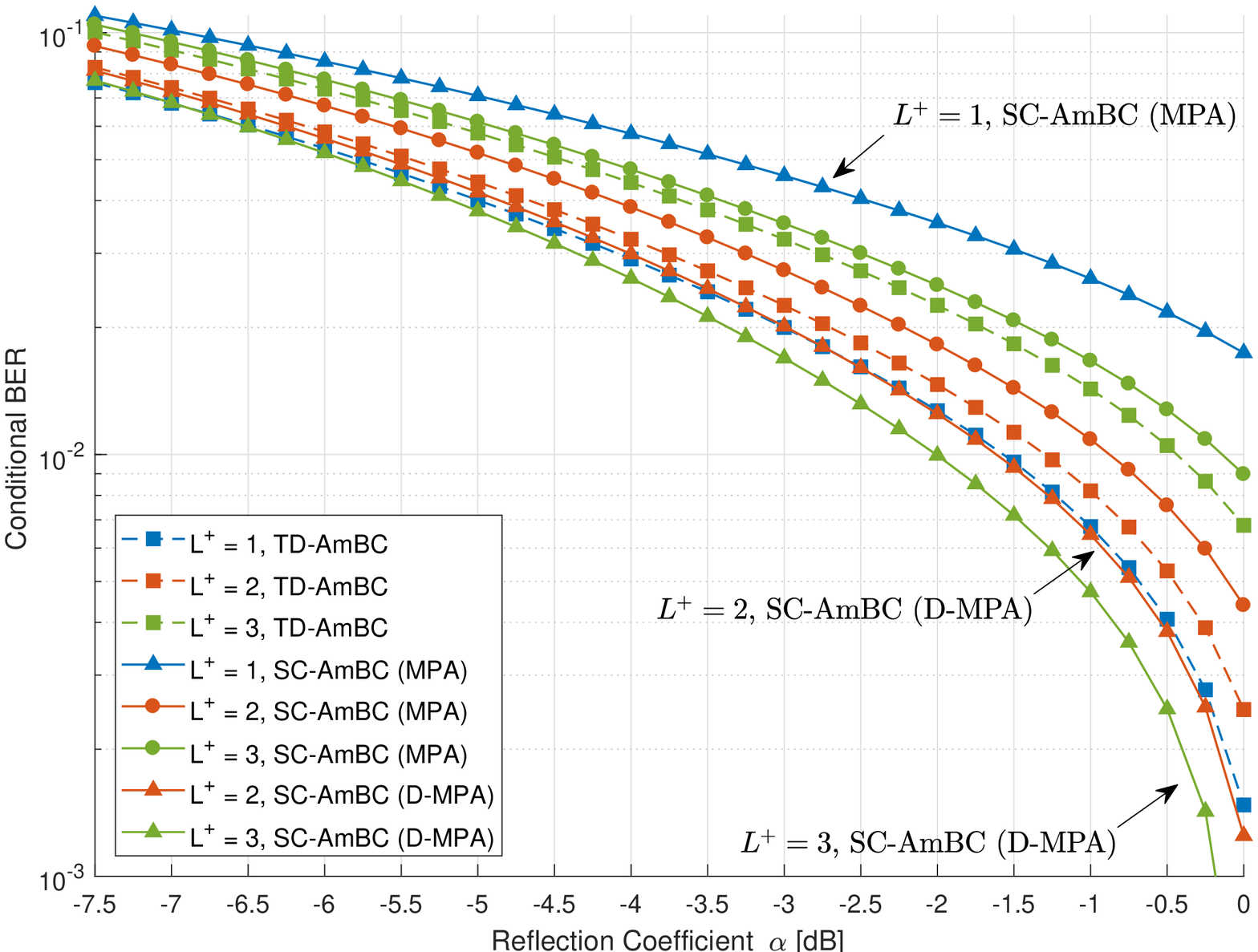}
(b) Conditional BER $1-p\left(\hat{\mathcal{I}} \big | \mathcal{E} \right)$}
\vspace{-0.0cm}
\caption{Effect of the number of the multipaths $L^{+}$ ($M=4$, $K = 4$, $\lambda=1.5$ for SC-AmBC).}
\label{fig:DBC}
\end{center}
\end{figure*}

\begin{figure*}[h]
\begin{center}
\parbox{7.5cm}
{\small \centering \includegraphics [width=7.5cm]{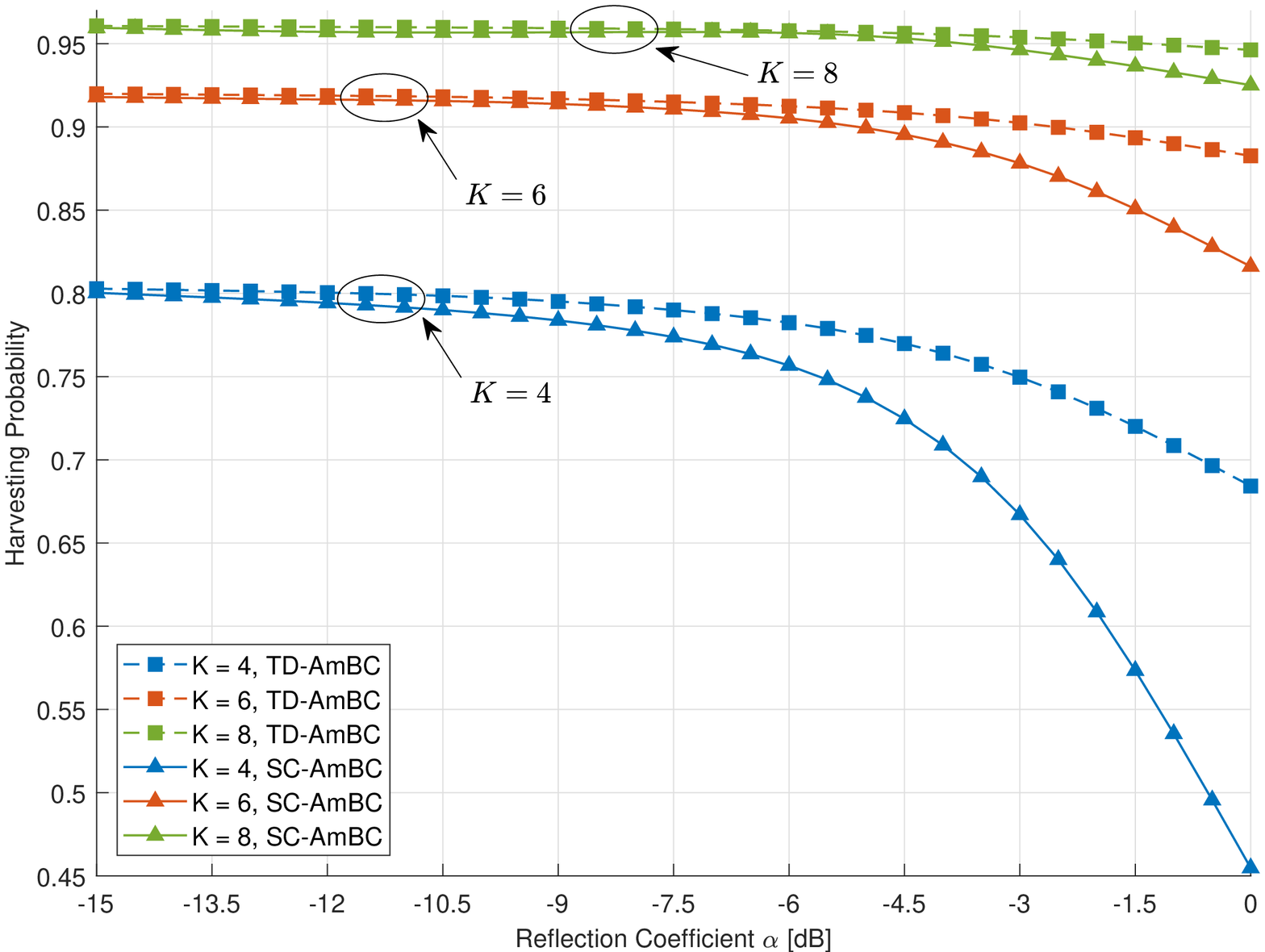} \\
(a) Harvesting probability $p\left(\mathcal{E} \right)$}
\quad
\parbox{7.5cm}
{\small \centering \includegraphics [width=7.5cm]{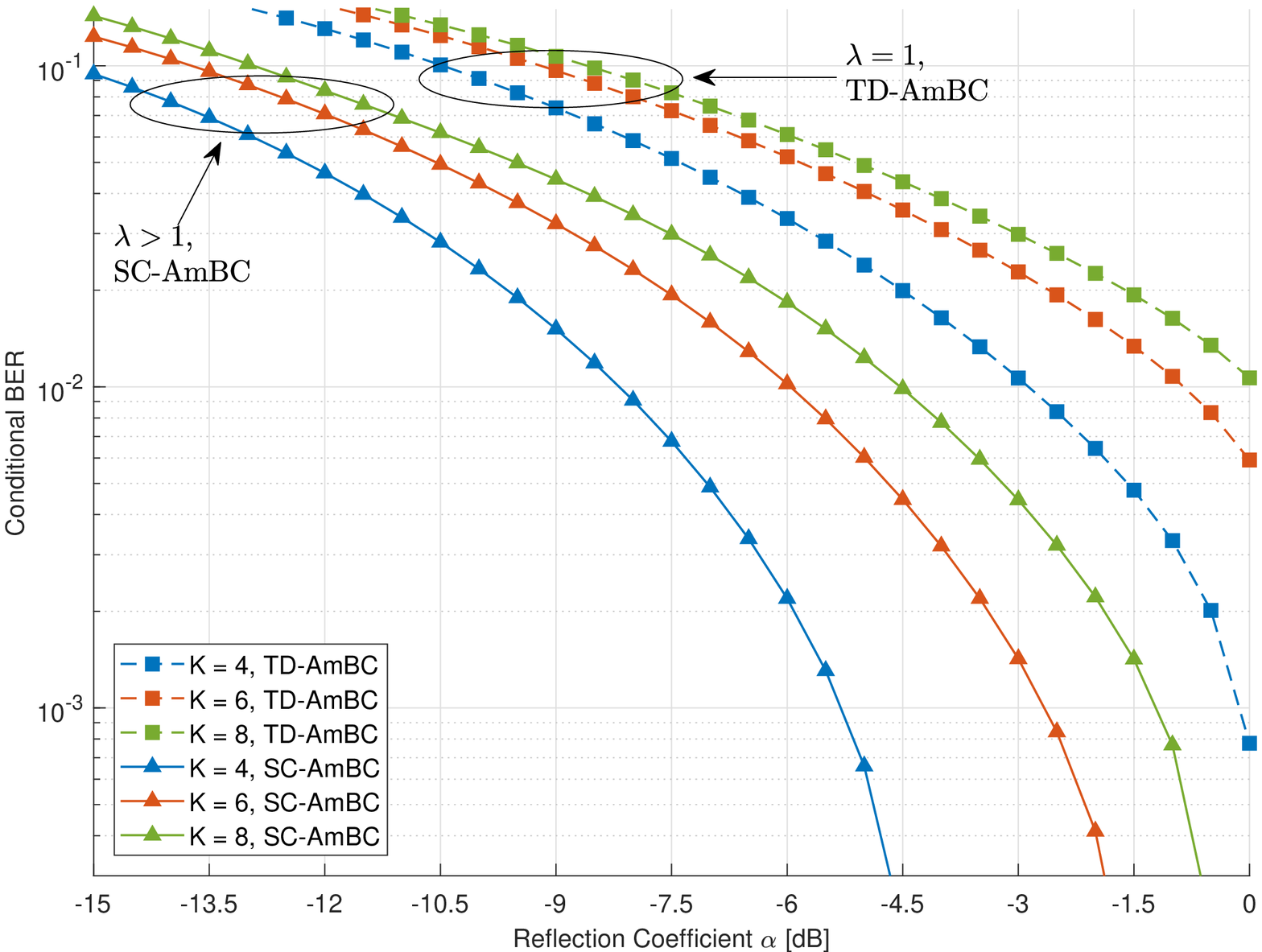}
(b) Conditional BER $1-p\left(\hat{\mathcal{I}} \big | \mathcal{E} \right)$}
\vspace{-0.0cm}
\caption{Effect of the number of the time slots $K$ ($M=2$, $L^{+} = 3$).}
\label{fig:OL}
\end{center}
\end{figure*}

\begin{figure*}[h]
\begin{center}
\parbox{7.5cm}
{\small \centering \includegraphics [width=7.5cm]{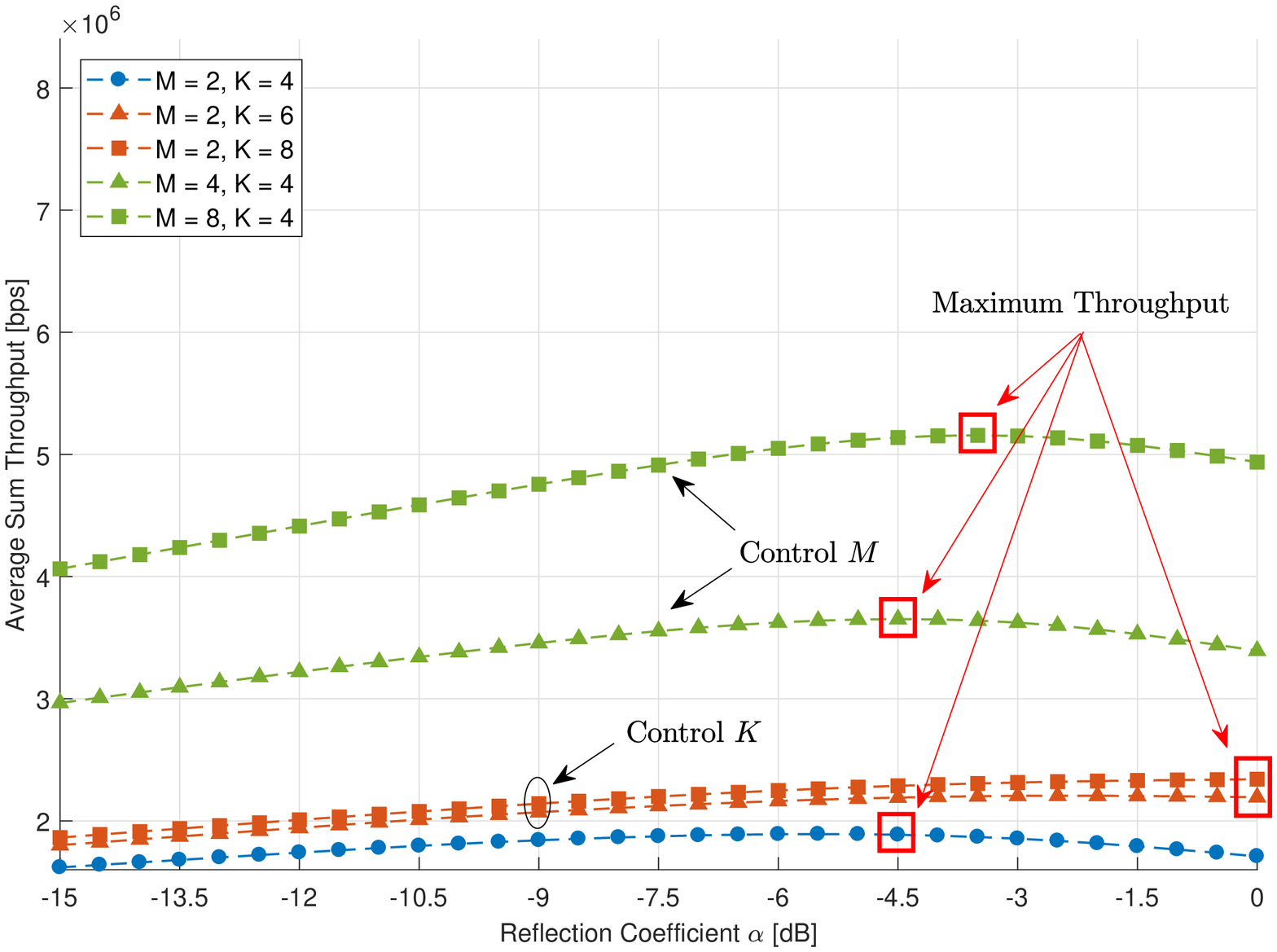} \\
(a) Average sum throughput $C$ (TD-AmBC)}
\quad
\parbox{7.5cm}
{\small \centering \includegraphics [width=7.5cm]{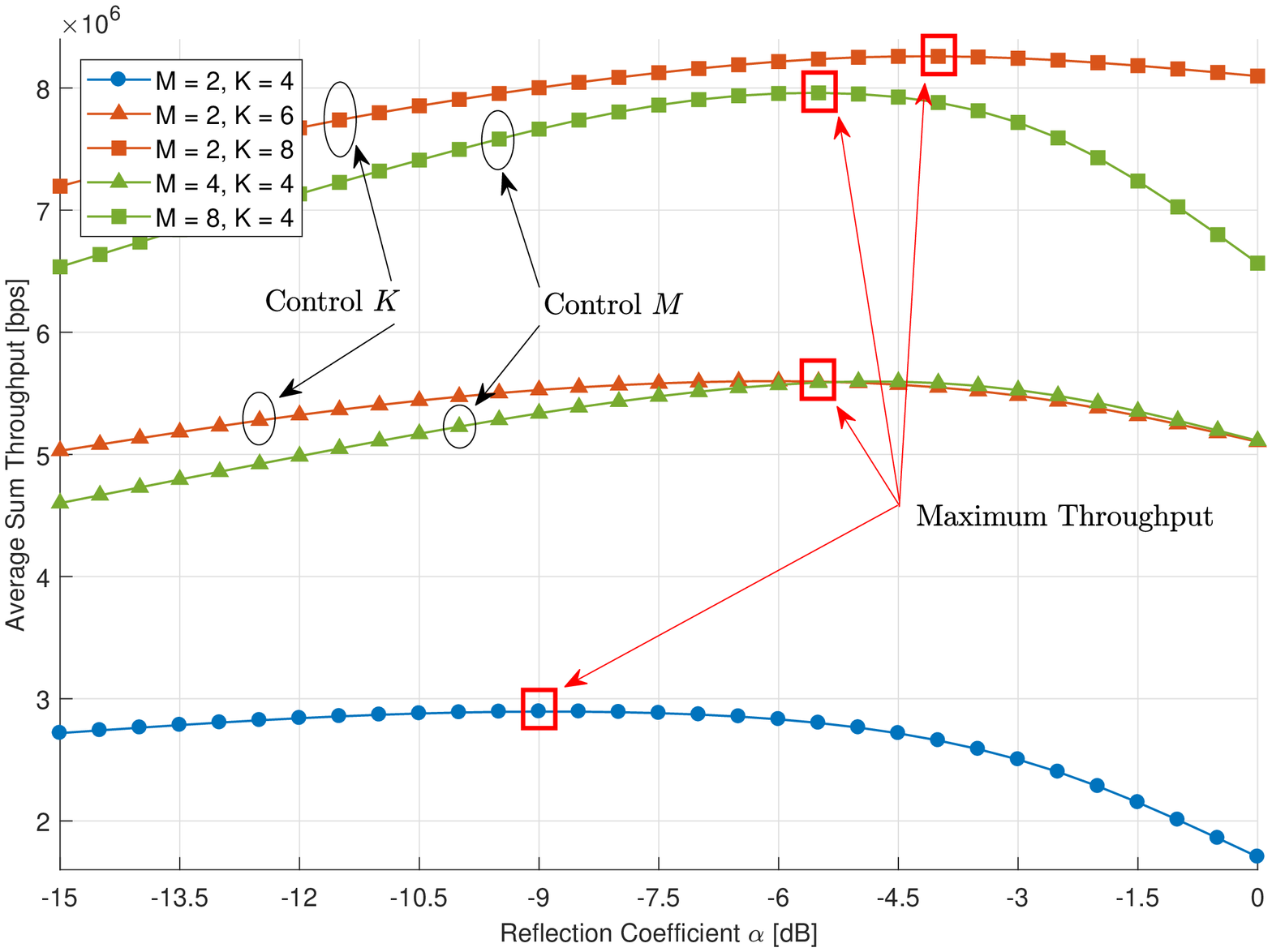}
(b) Average sum throughput $C$ (SC-AmBC)}
\vspace{-0.0cm}
\caption{Effect of the reflection coefficient $\alpha$ ($L^{+} = 3$).}
\label{fig:TH}
\end{center}
\end{figure*}

Fig.~\ref{fig:mod} shows the effect of the modulation order $M$ on the harvesting probability and the conditional BER.
In conventional TD-AmBC modulator, the harvesting probability is irrelevant to the value of $M$, since all the $M$-PSK symbols are on the circle of radius $\sqrt{\alpha}$ as shown in Fig.~\ref{fig:signal}. (a) with $\beta=0.25\alpha, \forall M$ in \eqref{eqn:NEPS}. 
Note that the reflection coefficient $\alpha$ in the horizontal axis is related to SNR per symbol rather than SNR per bit.
Consequently, as $M$ increases, conditional BER in TD-AmBC degrades as the SNR per bit is reduced by factor $\frac{1}{\log_2 M}$  \cite{wireless}. 

On the other hand, in SC-AmBC modulator, the harvesting probability is affected by the modulation order $M$ and the backscatter power $\sigma_b^2$ in \eqref{eqn:backpower}, which can be analyzed by the value of $\beta$ and the threshold $\theta$ in \eqref{eqn:EH-constraint}.
In 2-ary modulation, the normalized energy per symbol is given by $\beta = 0.5\alpha$, which is larger than those in TD-AmBC modulations. 
Larger value of $\beta$ increases the threshold value $\theta$ and eventually reduces the harvesting probability. 
Similarly, we can anticipate the performance degradation in 8-ary modulation with $\beta = 0.375\alpha$. 
However, we can observe no performance degradation especially for 4-ary modulation. 
In the 4-ary SC-AmBC, the duty cycle $D$ is doubled by using spreading sequences with factor $K_1=2$ and the backscatter power $\sigma_b^2$ is reduced to half by using zero-constellation point represented in Fig.~\ref{fig:signal}. (b). 
As a result, the non-orthogonal signaling with zero symbols (i.e., idle state) in SC-AmBC can compensate the harvesting probability without energy loss.

In terms of the BER performance in Fig.~\ref{fig:mod}. (b), all the SC-AmBC schemes outperform the TD-AmBC schemes.
Extension of signal space via sparse coding can enable 1.5X overloading as well as achieving significant diversity gains. 
Consequently, it is noted that for large $\alpha$, 4-ary SC-AmBC provides even lower BER than 2-ary TD-AmBC with same $\beta$. 
Although the average backscatter power is the same, SC-AmBC can utilize instantaneous channel fluctuations among different time slots, achieving modulation diversity \cite{mod-div}. 
Therefore, the proposed SC-AmBC can support small-form RF devices with $M$-ary modulation.

\subsection{Applicability to DBC}
Fig.~\ref{fig:DBC} shows the effect of the DBC for harvesting probability and conditional BER.
The harvesting probability increases as the number of multipath in forward channel, $L^{+}$ increases. 
The multipath diversity reinforces the incident power at tag antenna in \eqref{eqn:incident}, improving efficiency of ambient energy harvesting in both the TD-AmBC and the SC-AmBC. 
However, multipath is harmful to BER performance in conventional TD-AmBC. 
To resolve ISI caused by DBC, simply filtering ISI-affected samples in detector \cite{backfi, ofdm2} incurs significant loss in SNR as the number of multipath $L^{+}$ increases.
Moreover, this limits the operational range of backscatter applications such as inventory control in warehouse and grocery stores \cite{drone}. 
Thus, we can observe the tradeoff between the harvesting probability and the conditional BER in the TD-AmBC where ISI is not exploited.
  
On the other hand, the proposed SC-AmBC can fully exploit the ISI based on D-MPA and D-CEA.
Without these algorithms, we can partially exploit the ISI for performance improvement.
As shown in Fig.~\ref{fig:DBC}. (b), traditional MPA for one-way channel \cite{LDS, iter-Mary, code-project} shows worse BER than those of TD-AmBC, especially in DBC. 
However, if we add the weighted-sum of ISI for information calculation in \eqref{eqn:fk_ISI} using D-MPA, the conditional BER decreases as $L^{+}$ increases, which is in sharp contrast to the conventional TD-AmBC detectors and the baseline SC-AmBC with MPA. 
Therefore, the proposed D-MPA with D-CEA provides multipath diversity for both the energy harvesting and the information decoding by fully exploiting ISI in DBC.

\subsection{Applicability to NOMA}

Fig.~\ref{fig:OL} shows the effect of the number of time slots $K$ on the harvesting probability and the conditional BER. 
Due to high computational complexity for large $K$, we use $M=2$ for this scenario. 
As the number of time slots $K$ increases, the duty cycle decreases as $D = 1/K$ for TD-AmBC and $D=2/K$ for SC-AmBC. Consequently, backscatter power is reduced to $\beta = \frac{\alpha}{K}$ for TD-AmBC and $\beta = \frac{2\alpha}{K}$ for SC-AmBC, providing sufficient power to rectifier circuits in RF tags. 
Remarkably, increasing $K$ reduces the performance gap between the TD-AmBC and the SC-AmBC.
In this manner, increasing $K$ improves the harvesting probability by reducing the duty cycle $D$ in Fig.~\ref{fig:OL}. (a).

On the other hand, increasing $K$ degrades the conditional BER as shown in Fig.~\ref{fig:OL}. (b).
Since the backscatter power $\beta$ is reduced in both the TD-AmBC and the SC-AmBC in \eqref{eqn:NEPS}, the BER performance is degraded accordingly.
It is noteworthy that the TD-AmBC detectors have smaller backscatter power (i.e., $\beta = \frac{\alpha}{K}$) than those of the SC-AmBC (i.e., $\beta = \frac{2\alpha}{K}$), exhibiting the worse BER in Fig.~\ref{fig:OL}. (b).
More importantly, the sparsity, caused by small duty cycle is not utilized in those TD-AmBC detectors.
Those detectors can only support OMA with limited number of RF tags (i.e., $\lambda = 1$) and thus are not suitable for massive connectivity.
The proposed SC-AmBC can efficiently utilize the sparsity for reliable AmBC detection as well as for overloading multiple RF tags (i.e., $\lambda > 1$), supporting NOMA. Although MAI may degrade the BER performance, low-complexity MPA achieves near-optimal detection performance by the virtue of the sparsity \cite{scma1},
As a result, the proposed SC-AmBC can enable massive and reliable connectivity over the conventional TD-AmBC.

\subsection{Applicability to Energy Harvesting}

Fig.~\ref{fig:TH} shows the effect of the reflection coefficient $\alpha$ for the average sum throughput. 
The magnitude of reflection coefficient $\alpha$ balances the incident power arrived at tag antenna in \eqref{eqn:incident} between energy harvesting and information decoding. The energy-information tradeoff in AmBC determines the average sum throughput in various modulators and detectors \cite{ABCN-kaibin}.
However, in the conventional TD-AmBC, the throughput is only marginally improved as it does not exploit MAI and ISI by assuming orthogonality for the modulation and detection, as shown in Fig.~\ref{fig:TH}. (a).

The proposed SC-AmBC, on the other hand, is designed to exploit these interferences in non-orthogonal manner. 
Fig.~\ref{fig:TH}. (b) demonstrates the benefits of sparse coding in the throughput.
Compared to $5.15$ Mbps throughput in TD-AmBC, the maximum throughput of SC-AmBC is $8.26$ Mbps, achieving additional 60\% gain. 
In addition, the energy-information tradeoff can be flexibly adapted by controlling $K$, $M$, as well as $\alpha$ in SC-AmBC.
In conclusion, the proposed SC-AmBC is an energy-efficient solution for passive RF devices based on energy harvesting.

\section{Conclusions}\label{sec:conclusion}
In this paper, we have proposed AmBC modulator and detector based on sparse code to address low connectivity and channel fading issues inherent in AmBC.
The sparse code was employed at RF tags relying on ambient energy harvesting and supports non-orthogonal concurrent backscatter communications providing massive connectivity to Wi-Fi AP.
The sparse code using SC-BMA can be implemented to small-form factor RF devices and enables NOMA with $M$-ary modulation, achieving modulation diversity by extending the dimension of signal space.
For signal detection, the iterative D-MPA and the channel estimation based on D-CEA were proposed to overcome the DBC.
By exploiting MAI and ISI in modulation and detection, energy efficiency of AmBC can be enhanced, realizing massive connectivity for passive RF devices relying on energy harvesting.
Therefore, SC-AmBC can be applied to batteryless IoT including biosensors, wearables and smart homes where massive number of these RF devices are seamlessly integrated into existing wireless infrastructure.

\section*{Acknowledgment}
This work was supported by the National Research Foundation of Korea (NRF) grant funded by the Korean government (MSIT) (NRF-2014R1A5A1011478).

%
%
%
%
%




\end{document}